\documentclass[10pt,aps,prl,twocolumn,superscriptaddress,showkeys]{revtex4}

\usepackage{graphicx}
\usepackage{amsmath, amsthm, amssymb}
\usepackage{epsfig}
\usepackage{dcolumn}
\usepackage{rotating}
\usepackage{multirow}
\usepackage{epsfig}
\usepackage{color}
\usepackage{float}
\usepackage{setspace}

\begin{document}

\title{The optimal geometry of transportation networks}

\author{David Aldous}
\affiliation{Department of Statistics\\
367 Evans Hall $\#$ 3860\\
U.C. Berkeley CA 94720\\
\texttt{\small aldousdj@berkeley.edu}}
\author{Marc Barthelemy}
\affiliation{Institut de Physique Th\'eorique\\
CEA, IPhT, CNRS-URA 2306, F-91191 Gif-sur-Yvette, France}
\affiliation{CAMS (CNRS/EHESS) 54, boulevard Raspail\\
F-75006 Paris, France\\
\texttt{\small marc.barthelemy@ipht.fr}}

\date{\today}                                           

\begin{abstract}

  Motivated by the shape of transportation networks such as subways,
  we consider a distribution of points in the plane and ask for the
  network $G$ of given length $L$ that is optimal in a certain sense.
  In the general model, the optimality criterion is to minimize the
  average (over pairs of points chosen independently from the
  distribution) time to travel between the points, where a travel path
  consists of any line segments in the plane traversed at slow speed
  and any route within the subway network traversed at a faster
  speed.  Of major interest is how the shape of the optimal network
  changes as $L$ increases.  We first study the simplest variant of
  this problem where the optimization criterion is to minimize the
  average distance from a point to the network, and we provide some
  general arguments about the optimal networks.  As a second variant
  we consider the optimal network that minimizes the average travel
  time to a central destination, and discuss both analytically and
  numerically some simple shapes such as the star network, the ring or
  combinations of both these elements. Finally, we discuss numerically
  the general model where the network minimizes the average time
  between all pairs of points. For this case, we propose a scaling
  form for the average time that we verify numerically. We also show
  that in the medium-length regime, as $L$ increases, resources go
  preferentially to radial branches and that there is a sharp
  transition at a value $L_c$ where a loop appears.

\end{abstract}


\maketitle

\section{Introduction}

\begin{figure}[ht!]
\includegraphics[width=0.9\linewidth]{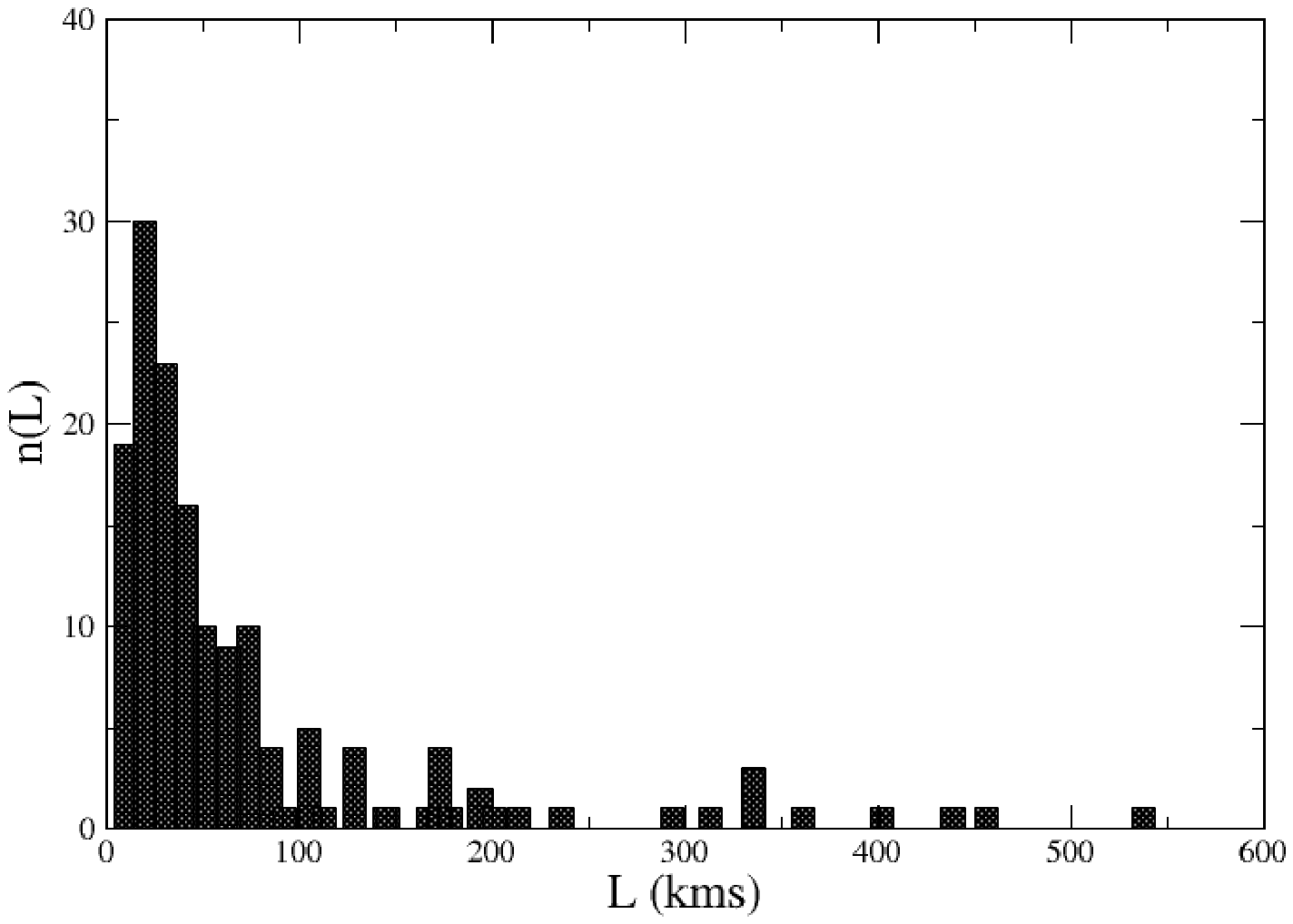}
\includegraphics[width=0.9\linewidth]{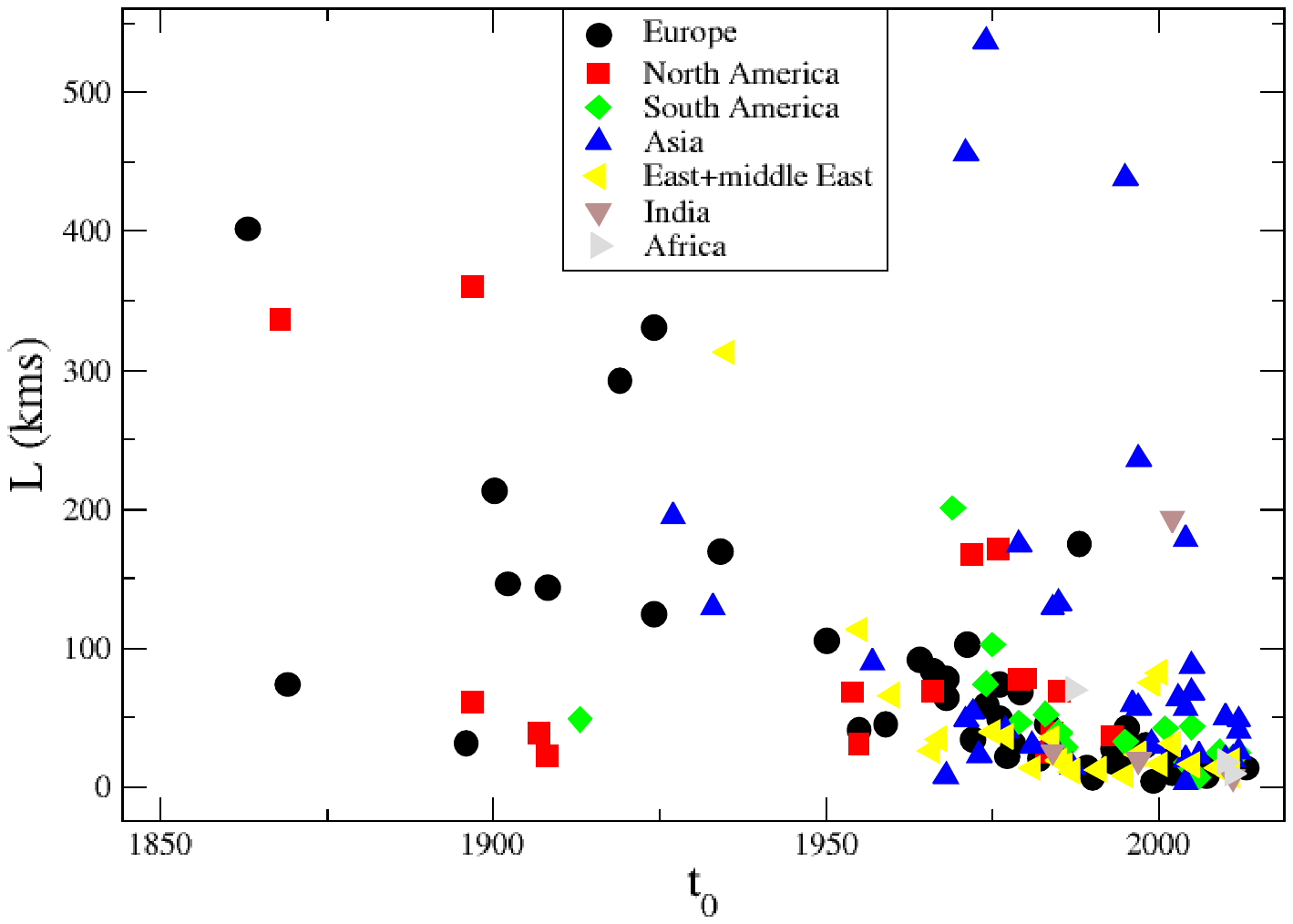}
\caption{(Top) Histogram of the total length of subway
  networks. (Bottom) Total length of the network versus first
  construction date. We grouped the networks according to broad
  regions.}
\label{fig:Ltot}
\end{figure}

Transportation networks evolve in time and their structure has been
studied in many contexts from street networks to railways and subways 
\cite{Xie:2011,Roth:2012,Barthelemy:2018,Bottinelli:2019}. The evolution of
transportation networks is also relevant in biological cases such as
the growth of slime mould \cite{Tero:2010} or for social insects
\cite{Latty:2011,Perna:2012,Ma:2013,Bottinelli:2015}. The specific
case of subways is particularly interesting (for network analysis of subways, see for example 
\cite{Latora:2002,Lee:2008,Derrible:2010,Derrible:2012,Roth:2012,Louf:2014}). In
most very large cities, 
a subway system has been built and later enlarged 
\cite{Roth:2012}, with current 
lengths $L$ varying from a few kilometers to a few hundred
kilometers. We observe that the length of subway networks is distributed over a broad range (see
Fig.~\ref{fig:Ltot}(top)).  Fig.~\ref{fig:Ltot}(bottom) also 
shows the total length versus the first construction date for most 
subway networks worldwide (the data is from various sources, see
\cite{Louf:2014} and references therein): the oldest networks are
mostly European and the largest and more recent ones can be found in Asia.

\begin{figure}[ht!]
\includegraphics[width=0.9\linewidth]{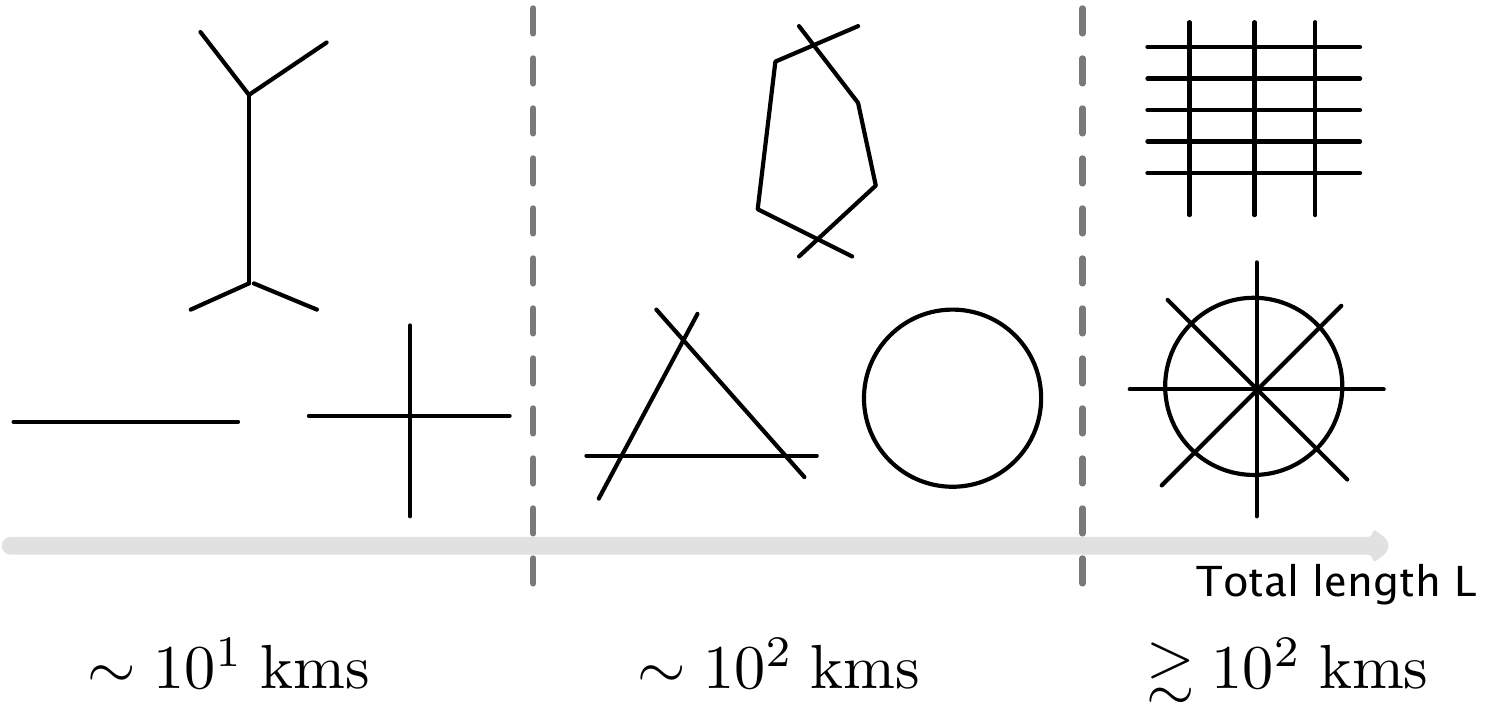}
\caption{Typical observed shapes when the length $L$ increases. For small
 $L$ we observe a line or a simple tree. For larger $L$ we
  observe the appearance of a loop and for much larger $L$ more
  complex shapes including a lattice like network or a superimposition
  of a ring and radial lines.}
\label{fig:typo}
\end{figure}

Concerning the geometry of these networks, as $L$ increases we observe
more complex shapes and an increase in the number of lines (see
Fig.~\ref{fig:typo} and also \cite{wiki}). Usually for small subways
($L$ of order a couple of 10\ kms) we observe a single line or a
simple tree (eg. a single line in the case of Baltimore, Haifa,
Helsinki, Hiroshima, Miami, Mumbai, Xiamen, \ldots ; or many radial
lines such as in Atlanta, Bangalore, Incheon, Kyoto, Philadelphia,
Rome, Sendai, Warsaw, Boston, Budapest, Buenos Aires, Chicago, Daegu,
Kiev, Los Angeles, Sapporo, Tehran, Vancouver, Washington DC). For
larger $L$ (of order 100\ kms), we typically observe the appearance of a
loop line, either in the form of a single ring (e.g.~ Glasgow) or
multiple lines with connection stations (Athens, Budapest, Lisbon,
Munich, Prague, S\~{a}o Paulo, St. Petersburg, Cairo, Chennai, Lille,
Marseille, Montreal, Nuremberg, Qingdao, Toronto). For larger networks
($L$ over 200 \ kms) we observe in general some more complex topological
structure (Berlin, Chongqing, Delhi, Guangzhou, Hong Kong, Mexico
City, Milan, Nanjing, New York, Osaka, Paris, Shenzhen, Taipei). For
the largest networks, convergence to a structure with a
well-connected central core and branches reaching out to suburbs has
been observed \cite{Roth:2012}.

In this paper we investigate the optimal structure of transportation
networks, as a function of length $L$, for several related notions of
{\em optimal} involving minimizing travel time. Real-world subway
networks have developed under many other factors, of course, rather
than resulting from the optimization of some simple quantity, but
optimal structures provide interesting benchmarks for comparison with
real-world networks.

There has been extensive study of optimal networks over a given set of
nodes (such as the minimum spanning tree \cite{Graham:1985}, or other
optimal trees \cite{Barthelemy:2006}).  Some such problems allow extra
chosen nodes, for example the Steiner tree problem \cite{Hwang:1992},
or geometric location problems in which $n$ given demand points are to
be matched with $p$ chosen supply points \cite{Megiddo:1984}.  At
another extreme is the much-studied Monge-Kantorovich mass
transportation problem \cite{Rachev:1998}, involving matching points
from one distribution with points from another distribution.  Our
setting is fundamentally different, in that what we are given is just
the density of start/end points on the plane.  A network is
intrinsically one-dimensional, in the sense of being a collection of
(maybe curved) lines embedded in the plane.  In a sense we are
studying a coupling between a given distribution over points in the
continuum and a network of our choice constrained only by length and
connectedness.  Some simpler problems of this type have been addressed
previously. For instance, the problem of the quickest access between
an area and a given point was discussed in
\cite{Bejan:1996,Bejan:1998}. More recently, the impact of the shape
of the city and a single subway line was discussed in
\cite{Gettrick:2019}. Algorithmic aspects of network design questions
similar to ours have been studied within computational geometry
(e.g. \cite{Okabe:1992} chapter 9) and ``location science"
(e.g. \cite{Laporte:2015} and references therein).  But our specific
question -- optimal network topologies as a function of population
distribution and network length -- has apparently not been explicitly
addressed.

\section{The model and the main question}

Here we define precisely the general model we have in mind, and the
three different variants that we will in fact study.  Our model makes
sense for an arbitrary ``population density" $\rho( \ )$ on the plane,
but we will study mostly the isotropic (rotation-invariant around the
origin) case, in particular the (standard) Gaussian density
\[ \rho(x,y) =(2 \pi)^{-1} \exp(-r^2/2), \ r^2 = x^2 + y^2 \] and the
uniform distribution on a disc.  The population density is of
individuals who wish to reach as quickly as possible other points in
the system (for simplicity we do not distinguish densities of
residences and workplaces, for instance). Continuing with the subway
interpretation, we assume that one can move anywhere in the plane at
speed $1$, and one can move within the network at speed $S > 1$ (this
quantity can therefore be seen as the ratio $S=v_s/v_w$ between walking
and subway velocities). Note that we envisage each position on the
network as a ``station" where the subway is accessible -- the relative
efficiency of two networks would be little affected by the
incorporation of discrete stations into a model.

The general problem is the following one.  For any pair of points
$(z_1,z_2)$ in the plane, there is a minimum (over all possible
routes) time $\tau(z_1,z_2)$ to journey from $z_1$ to $z_2$, and so
the average journey time is
\[ \overline{\tau} = \int \int \tau(z_1,z_2) \rho(z_1) \rho(z_2) \
  dz_1 dz_2 .\] This depends on the network.  For a given $L$ there is
some optimal network of length $L$ (this network also depends on the
speed ratio $S$ and on $\rho$).  We study the shapes and average
journey times for such optimal networks.

This ``general model" is very simplistic -- as a next step, a
companion paper \cite{AB2} studies an extended model including a
waiting time $W$ whenever we take the subway or connect from one line
to another -- but nevertheless seems analytically intractable.  So in
fact we will consider three simpler variants.

First, we will consider the problem of minimizing the average (over
starting points from the given distribution) distance to the network,
that is to the closest point in the network.  
Note here
that we don't compute the average time between all pairs of points but
just consider the access time to the network from each point.  The
second variant that we consider is the problem of minimizing the
journey time from the points to a single destination, which we may
take to be the origin $O$.  In the third variant, we engage the
general issue of routes between arbitrary points which typically (but
not always) involve entering and exiting the subway network, but now
require these entrances and exits to be the closest positions to the
starting and ending points, rather than the time-minimizing positions.

Except in asymptotic results (e.g. equation (\ref{eq:asymp2})) we do
not have exact formulas involving optimal networks. Instead we consider a range of simple network shapes, 
 allowing us to investigate the possible shape of optimal networks. 

\section{A first simplification: optimal placement}

Here we consider the simplest variant, in which we seek the network
(of given length $L$) that minimizes the average distance from a point
to the network.  This is almost the same as the $S=\infty$ case of the
general model, because the journey time between two points would be
the sum of the two distances to the network, except that in the
general model the shortest route might not use the subway network at
all.  Intuitively an optimal network must come close to most points of
the distribution.  Although superficially similar to the notion of
space-filling curves \cite{Sagan:1994}, the latter are fractal curves
whereas our networks (having finite length) cannot have fractal
curves.

\subsection{Some rigorous results}

Here we outline some rigorous results for this variant model, with
details to be given in the companion paper \cite{AB2} .

Observe that given a straight line segment, the area within a small
distance $\varepsilon$ from the line is $2\varepsilon$ per unit
length.  So in a network of length $L$, the total area within that
distance from the network is at most $2\varepsilon\times L$, and is
reduced from that value by the presence of curved lines and
intersections.  By extending that argument one can prove \cite{AB2}
that the optimal network is always a tree (or a single curve, which is
a special case of a tree).  For a non-isotropic density $\rho( \ )$
the optimal network may not be a single line, but we conjecture that
for isotropic densities decreasing in $r$ the optimal network is
always a single curve.

Although $L \to \infty$ asymptotics are hardly realistic in the
context of subway networks, the same questions might arise in some
quite different context, so it seems worth recording the explicit
result for asymptotics.  In the $L \to \infty$ limit, the optimal
network density (i.e. the edge length per unit area of the network)
near point $z$ should be of the form $L\phi(\rho(z))$ for some
increasing function $\phi$.  By scaling, the average distance from a
typical point near $z$ to the network should be $c_0/(L\phi(\rho(z)))$
for some constant $c_0$.  So the overall average distance to the
network is
 \begin{align}
\overline{d}(L)=\frac{c_0}{L} \int\frac{1}{\phi(\rho(z))} \rho(z)  dz .
\label{eq:dL}
\end{align}
The total length constraint implies that
\begin{align}
\int \phi(\rho(z))\ dz= 1 .
\label{eq:con}
\end{align}
A standard Lagrange multiplier argument shows that the integral in
(\ref{eq:dL}) is minimized, over functions $\phi$ under constraint
(\ref{eq:con}), by a function of the form $\phi(\rho) = a \rho^{1/2}$,
and then (\ref{eq:con}) and (\ref{eq:dL}) combine to show
 \begin{align}
\overline{d}(L)_{opt}=
\frac{c_0}{L} \left(\int\rho^{1/2}(z) \ dz\right)^2 .
\label{eq:asymp}
\end{align}
Finally, the constant $c_0$ can be re-interpreted as the minimum
average-distance-to-network in the context of networks on the infinite
plane with network density $ = 1$.  From our initial ``area within a
small distance $\varepsilon$" observation, the optimal network in the
infinite context consists simply of parallel lines spaced one unit
apart, for which $c_0 = 1/4$.  So this analytic argument shows
  \begin{align}
\overline{d}(L)_{opt} \sim 
\frac{1}{4L} \left(\int\rho^{1/2}(z) \ dz\right)^2 \mbox{ as } L \to \infty .
\label{eq:asymp2}
\end{align}
 
This result makes no assumption about the underlying density $\rho$.
In the Gaussian case, the integral in (\ref{eq:asymp2}) equals
$\sqrt{8\pi}$ and so $\overline{d}(L)_{opt} \sim \frac{2 \pi}{L}$.

\begin{figure}[ht!]
\includegraphics[width=0.85\linewidth]{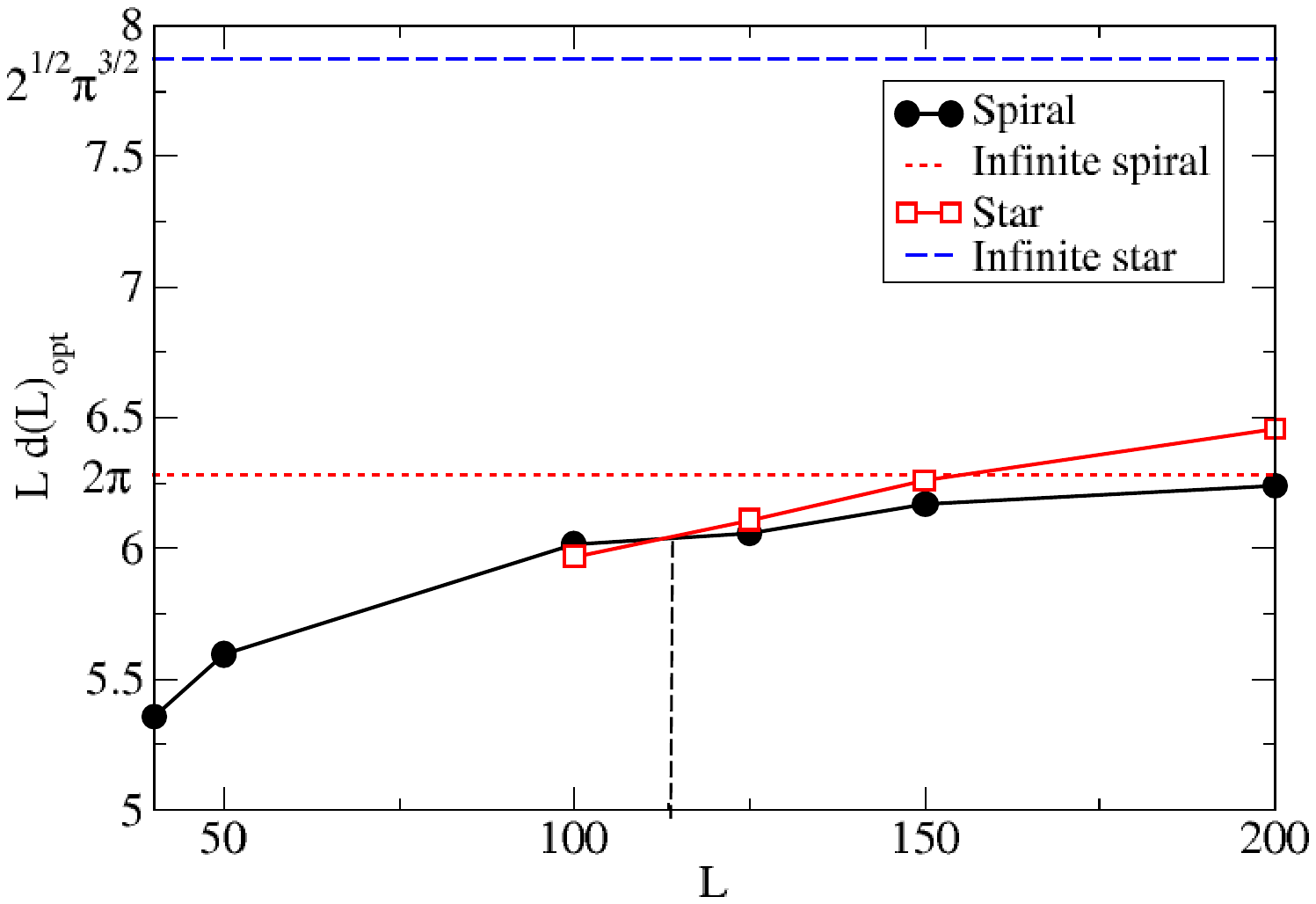}
\caption{Rescaled average distance $L\overline{d}(L)_{opt}$ versus $L$ for the spiral and the
  star network. For the spiral, we observe a slow
  convergence towards the theoretical limit $2\pi$ for gaussian
  disorder. The convergence is even slower for the star network which
  converges to $2^{1/2}\pi^{3/2}$. For $L\approx 110$ (dashed vertical
line) the spiral outperforms the star network.}
\label{fig:asymp}
\end{figure}

As explained in \cite{AB2}, what this argument actually shows is that
a sequence of networks is asymptotically optimal as $L \to \infty$ if
and only if the rescaled local pattern around a typical position $z$
consists of asymptotically parallel lines with spacing proportional to
$\rho^{-1/2}(z)$, but the orientations can depend arbitrarily on $z$.
Visualize a fingerprint.  For an isotropic density we can arrange such
a network to be a spiral.  This enables us to check the Gaussian
prediction numerically.  Consider a spiral of length $L$ starting at
some point $(a_L,0)$ and with rings at radius $r$ separated by
$b_L \exp(r^2/4)$, and then optimize over $(a_L, b_L)$.  Numerically we
find slow convergence toward this limit behavior, shown in
Fig.~\ref{fig:asymp}.




\subsection{Numerical study: different shapes}

Unfortunately the asymptotics above  say
nothing about the actual shapes of the optimal
networks for more realistic smaller values of $L$. 
Intuitively, we expect that for  very small $L$ the optimal network 
is just a line segment centered on the origin. As $L$ increases
we  expect a smooth transition from the line segment 
to a slightly-curved path, a ``C-shaped" curve.
But then how the optimal shapes transition  to a tight spiral path (presumed optimal for very large $L$) 
is not {\em a priori} easy to guess.

We tested 8 shapes numerically for the standard Gaussian distribution. 
In each shape the length is $L$, and we optimized over any free parameters 
(such as $s$ in the $2 \times 2$ grid). 

\begin{figure}[ht!]
\includegraphics[width=0.8\linewidth]{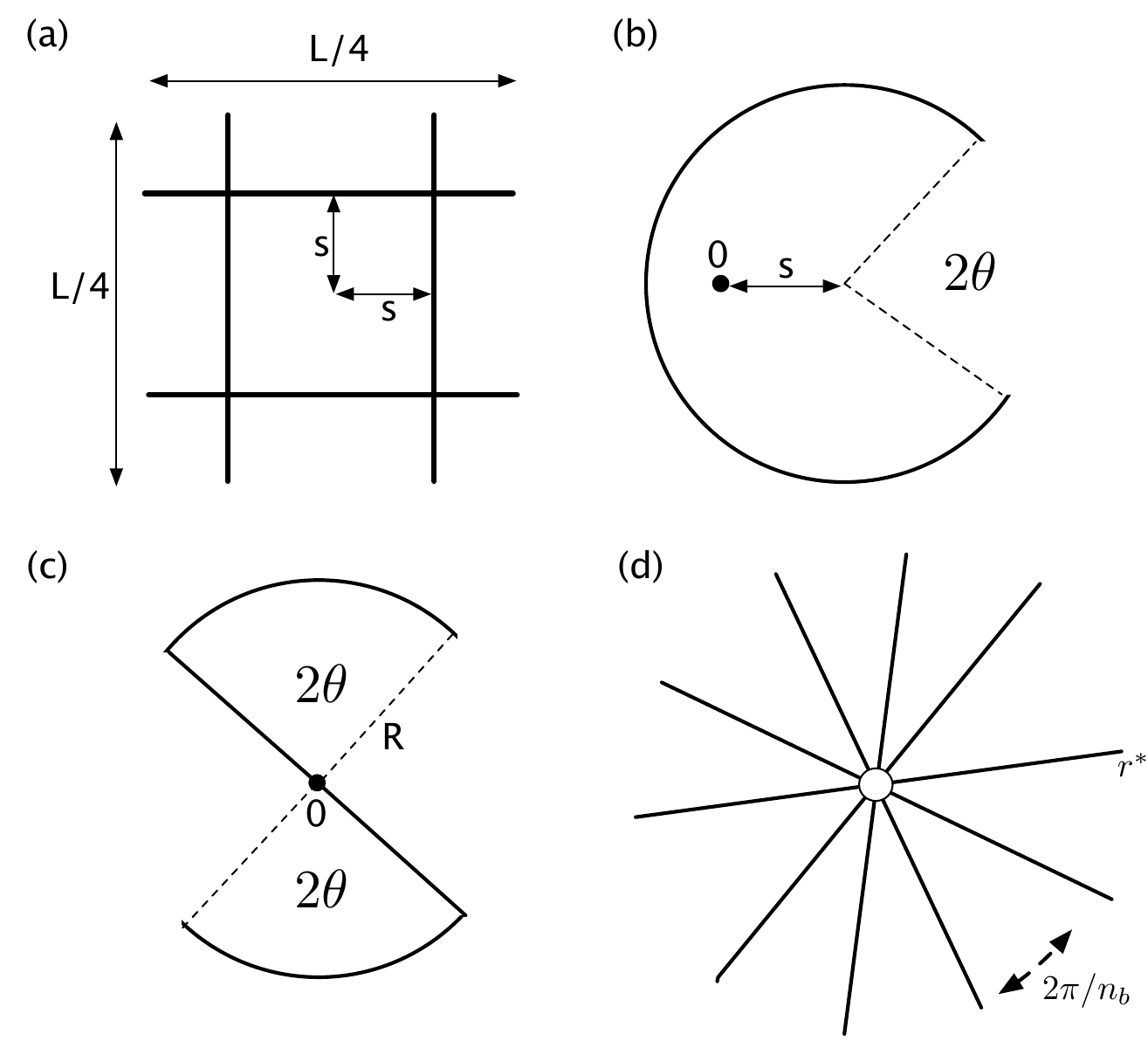}
\caption{Main shapes studied here (in addition to the line and the
  spiral). (a) The ``hashtag'' of $2\times 2$ grid with parameter
  $s$. (b) The ``C-shape'' with paramerers $s$ and $\theta$. (c) The
  ``S-shape'' with parameter $\theta$. (d) The star network with $n_b$
  branches of size $r^*$.}
\label{fig:mainshapes}
\end{figure}

\begin{itemize}
\item The line segment $[-L/2,L/2]$.
\item The ``cross" (two length $L/2$ lines crossing at the origin).
\item The ``hashtag" or ``$2 \times 2$ grid" (Fig.~\ref{fig:mainshapes}(a)).
\item The ``ring" (circle centered on the origin).
\item The ``C-shape" (off-centered partial circle, with arc length $2 \theta$ removed, see Fig.~\ref{fig:mainshapes}(b)).
\item The ``S-shape":  two arcs of circle of radius $R$ and of
  angle $2\theta$, connected by a straight line of length $2R$ (see Fig.~\ref{fig:mainshapes}(c)).
\item The ``star" with $n_b$ branches of length $r^*$ (so $r^* n_b=L$,
  see Fig.~\ref{fig:mainshapes}(d)). 
 \item The (Archimedean) ``spiral", $r = a\theta + b$. 
\end{itemize}

Recall that the optimal (over all shapes) shape is always a path or
tree, so the $2 \times 2$ grid or ring can never be overall optimal.
Note also that (for any isotropic distribution) the optimal ring has
radius equal to the median of the radial component of the underlying
distribution, in our Gaussian case $\sqrt{2 \log 2} \approx 1.18$.

\begin{figure}[ht!]
\includegraphics[width=0.9\linewidth]{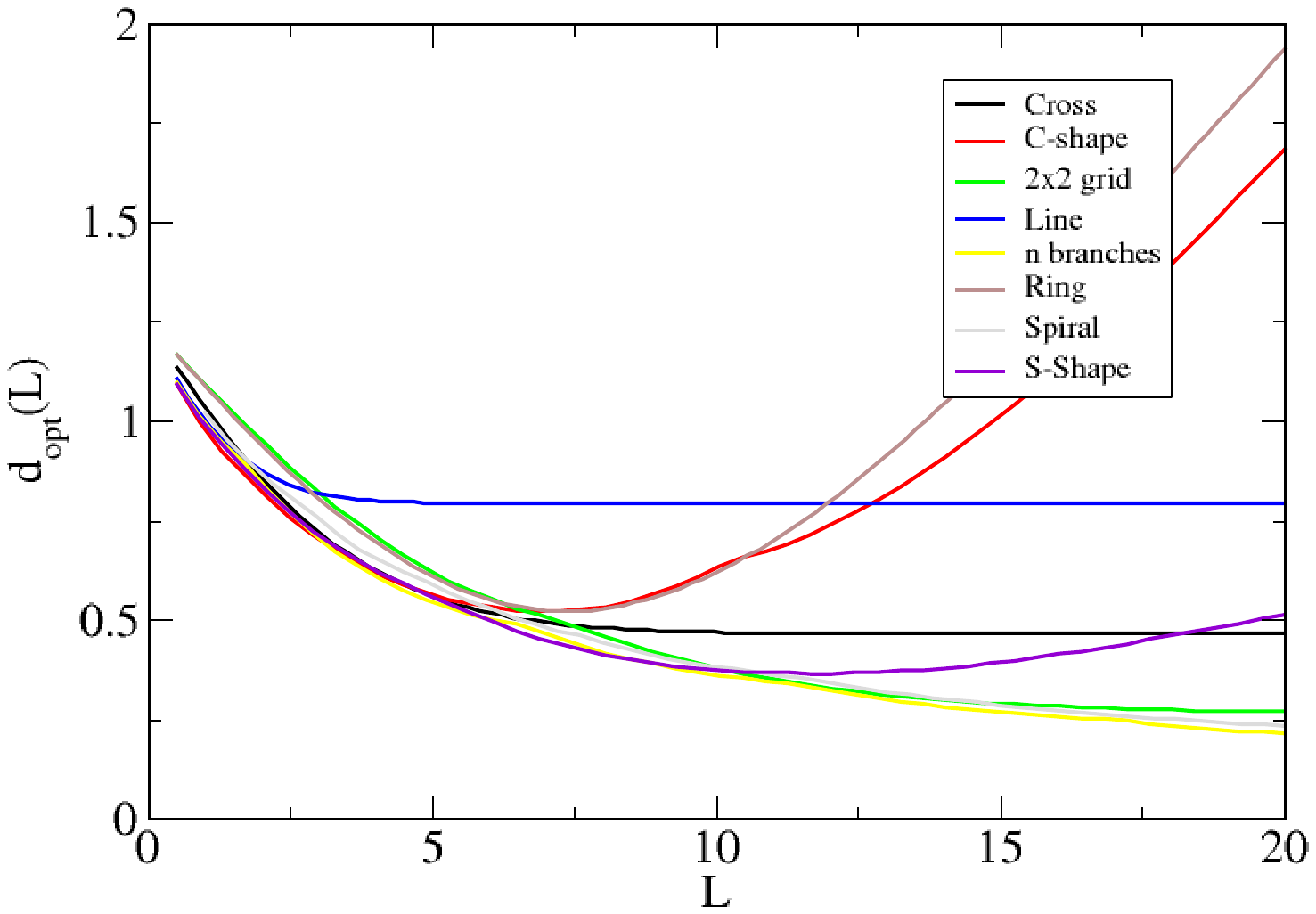}
\caption{Average distance to the network  versus L for various
  shapes.}
\label{fig:allshapes}
\end{figure}

We simulated these different shapes in the Gaussian disorder case, and for each
value of $L$ we optimize over the parameters defining the different
shapes. We show the results for these various shapes in
Fig.~\ref{fig:allshapes}.

The following general picture emerges:

\begin{itemize}
\item{} For small $L\lesssim 3.2$ the optimal shape is the C-shape.
  We note that for this shape the optimal $\theta$ decreases with $L$:
  for $L\approx 6$ we have $s=0$ and for $L\approx 8.0$ the optimal
  $\theta=0$. When $\theta=0$ and $s=0$ we then recover the ring
  result (see Fig.~\ref{fig:cshape}).
\begin{figure}[ht!]
\includegraphics[width=0.9\linewidth]{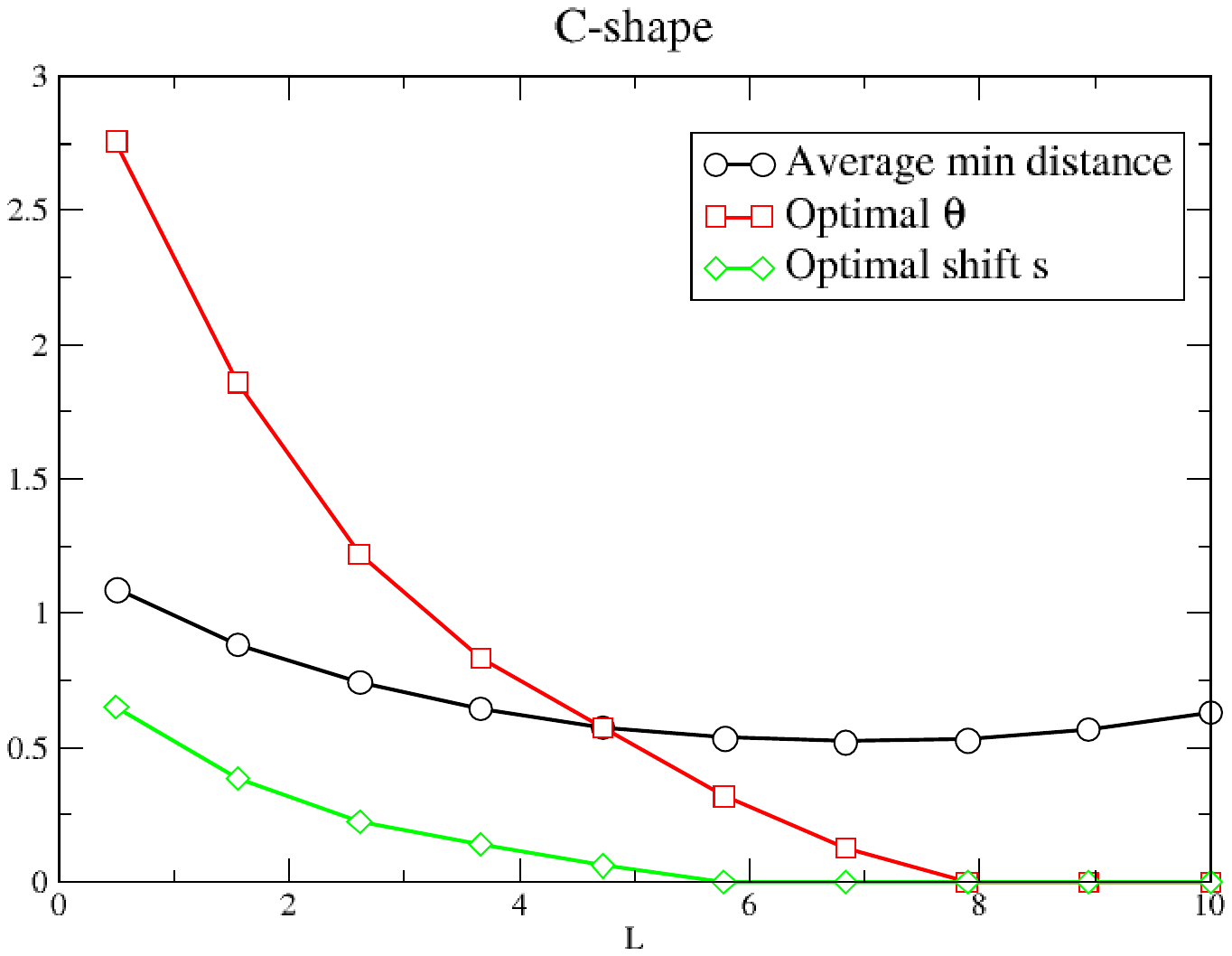}
\caption{Study of the C-shape with parameter $s$ and $\theta$.}
\label{fig:cshape}
\end{figure}

\item{} For $3.2\lesssim L\lesssim 5.8$, the cross (star network with 4 branches)
is optimal.

\item{} For $5.8\lesssim L\lesssim 8.7$, the S-shape is optimal. We
  note that for this shape, as $L$ increases, the optimal angle $\theta$ increases
  from $0$ to $\pi/2$, but with a jump at $L\approx 5$ (see
  Fig. ~\ref{fig:sshape}).
\begin{figure}[ht!]
\includegraphics[width=0.7\linewidth]{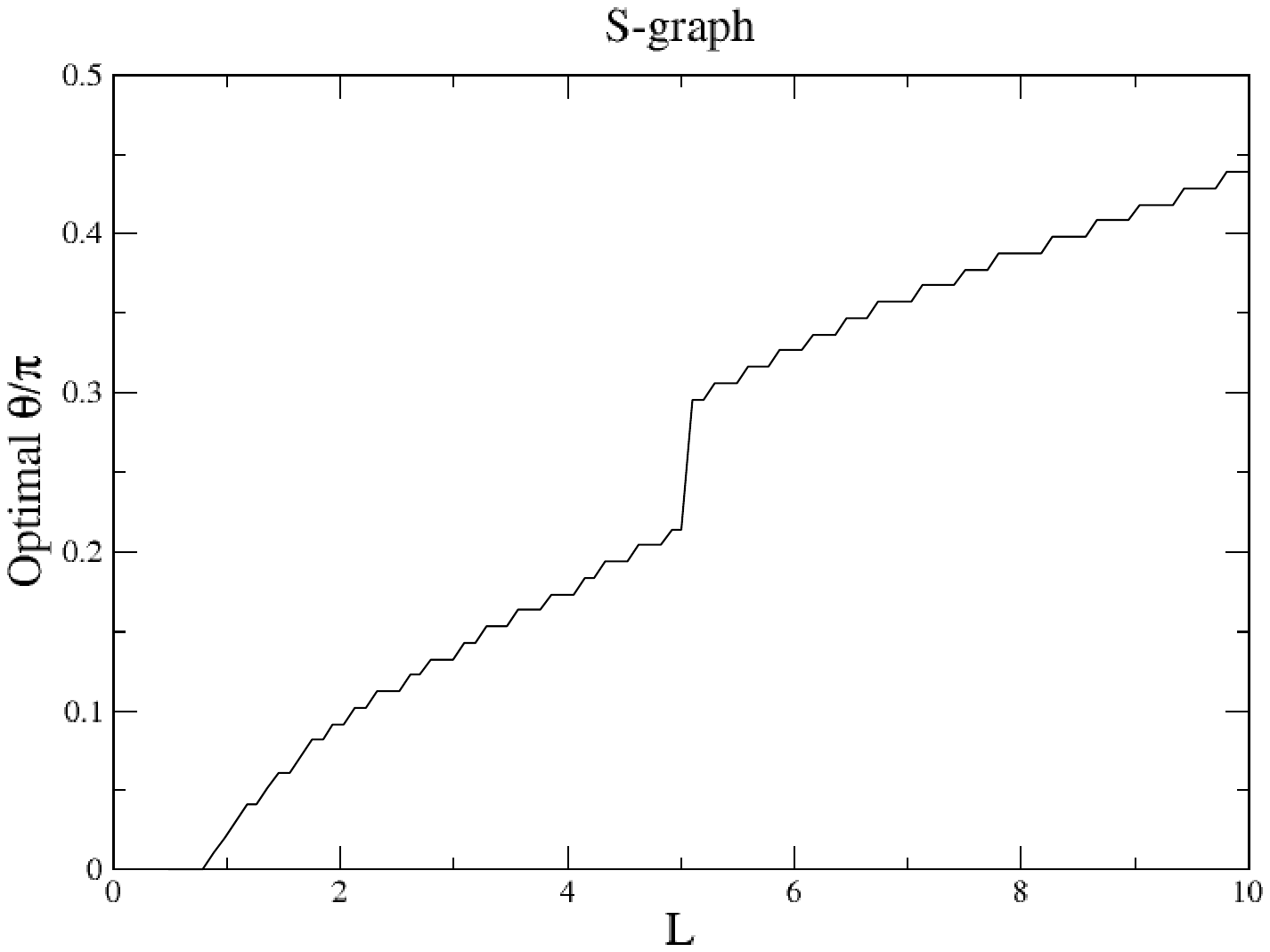}
\caption{Study of the S-shape: evolution of the optimal angle $\theta$
  versus $L$.}
\label{fig:sshape}
\end{figure}

\item{} For $L\gtrsim 8.7$ the star network with $n_b$ branches is the
  optimal shape. The number of branches is roughly increasing with
  $L$: $n_b\sim pL+q$ with $p\approx 0.4$. 
  
\item{} The simple Archimedean spiral was slightly less efficient than
  the star network over the range of $L$ considered above: For $L=20$,
  the average time is $\overline{d}_{opt}\approx 0.215$ for the star
  network, while for the spiral we have $\overline{d}_{opt}\approx 0.233$. 

  \end{itemize}

\subsubsection{Qualitative discussion}

In the examples above, the {\em shapes} were not adapted specifically
to the Gaussian model (although the numerical parameters were
optimized) and so are shapes one might consider for other isotropic
distributions.  Recall that our previous analysis of the
$L \to \infty$ behavior found optimal networks to be spiral-like in a
specific distribution-dependent way (near a point $z$, the rings are
separated by distance proportional to $(\rho(z))^{-1/2}$).
For comparison, it is
straightforward to show that the asymptotic behavior of the optimal
star shape is $ \overline{d}(L)_{opt} \sim 2^{1/2}\pi^{3/2} /L$, and
numerical results are shown in the Fig.~\ref{fig:asymp}. By comparing
with the spiral, we estimate that the value at
which such spiral networks out-perform star networks in the Gaussian
model is around $L = 110$.

\begin{figure}[ht!]
\includegraphics[width=0.45\linewidth]{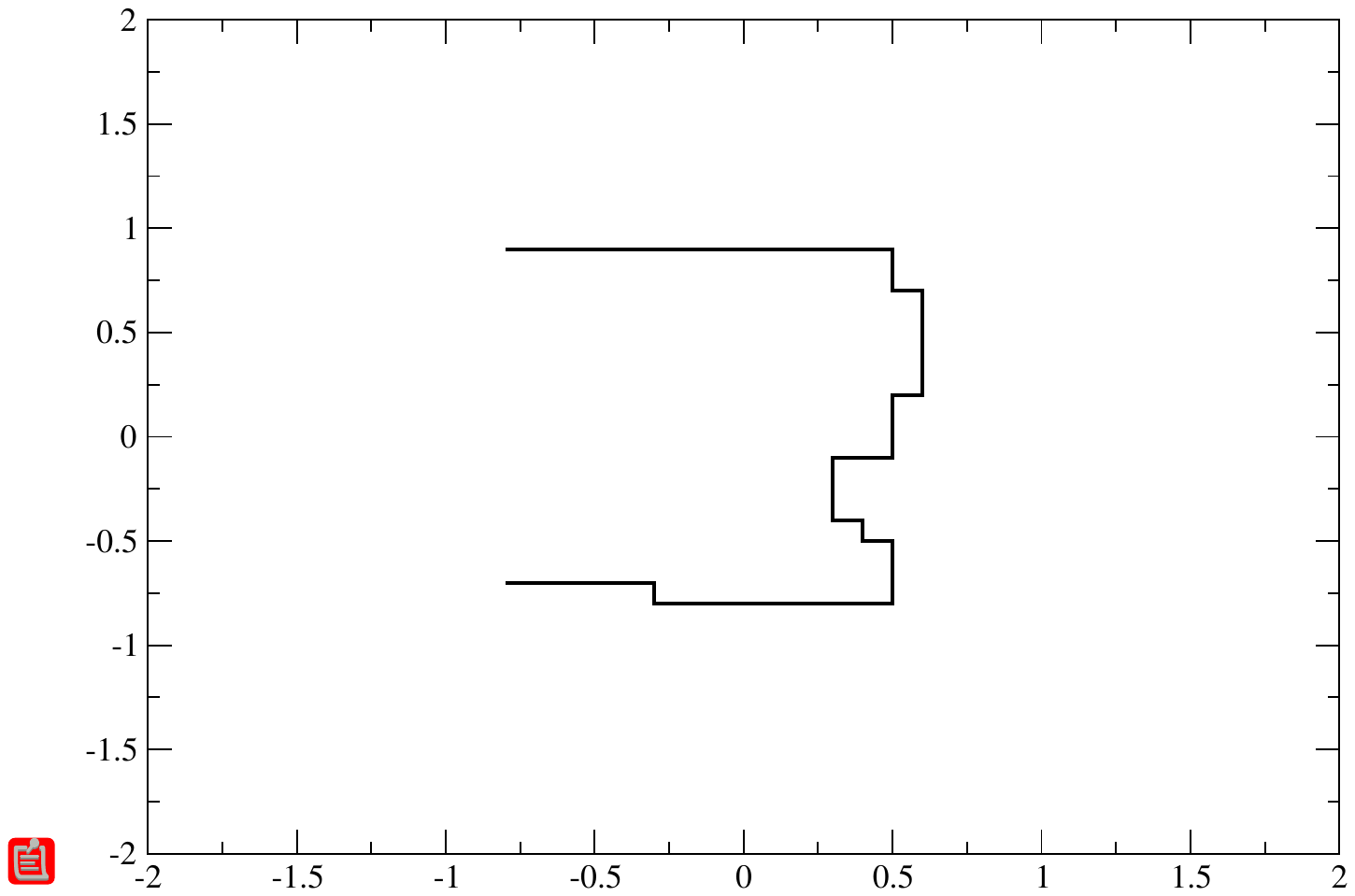}
\includegraphics[width=0.45\linewidth]{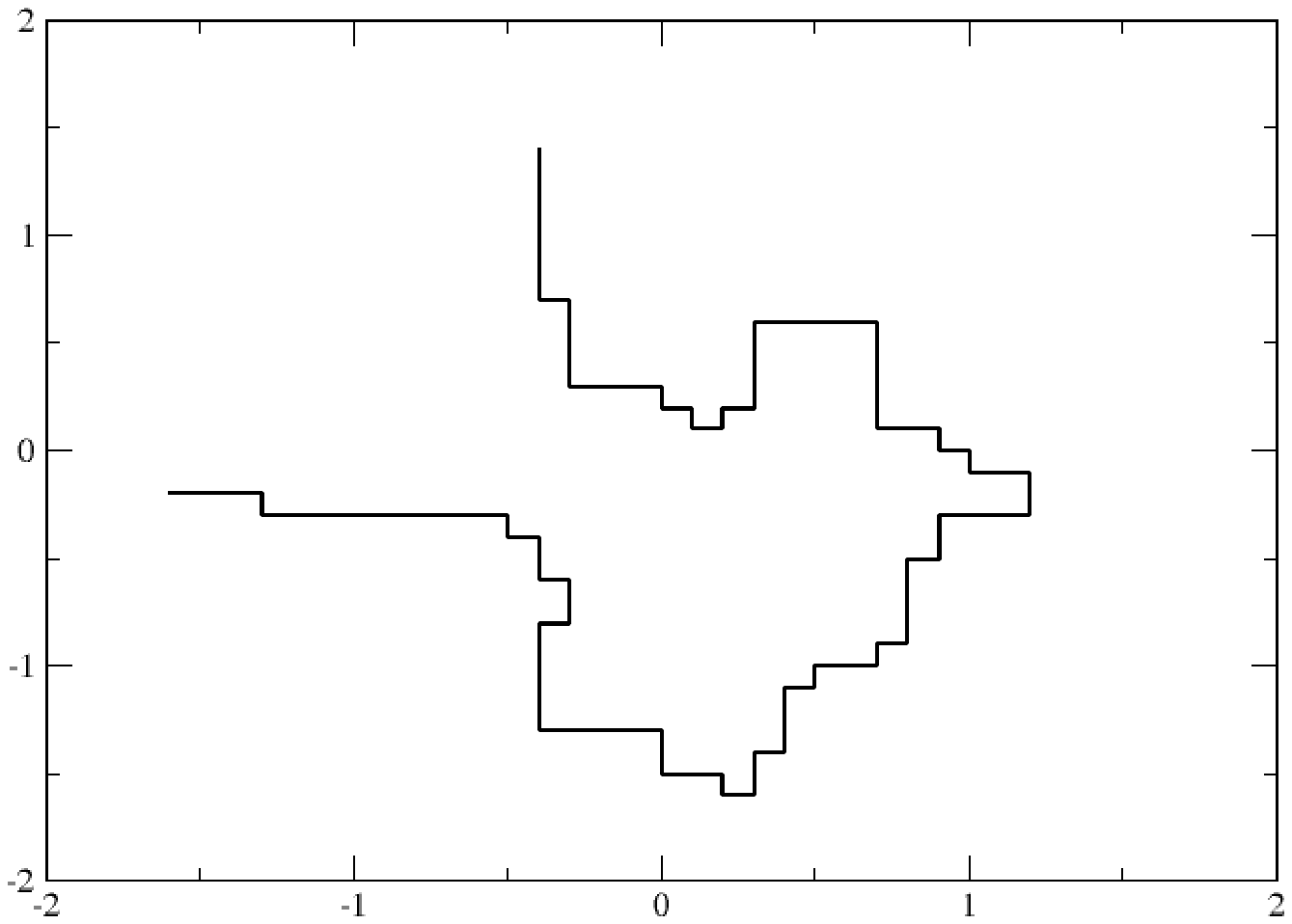}\\
\includegraphics[width=0.45\linewidth]{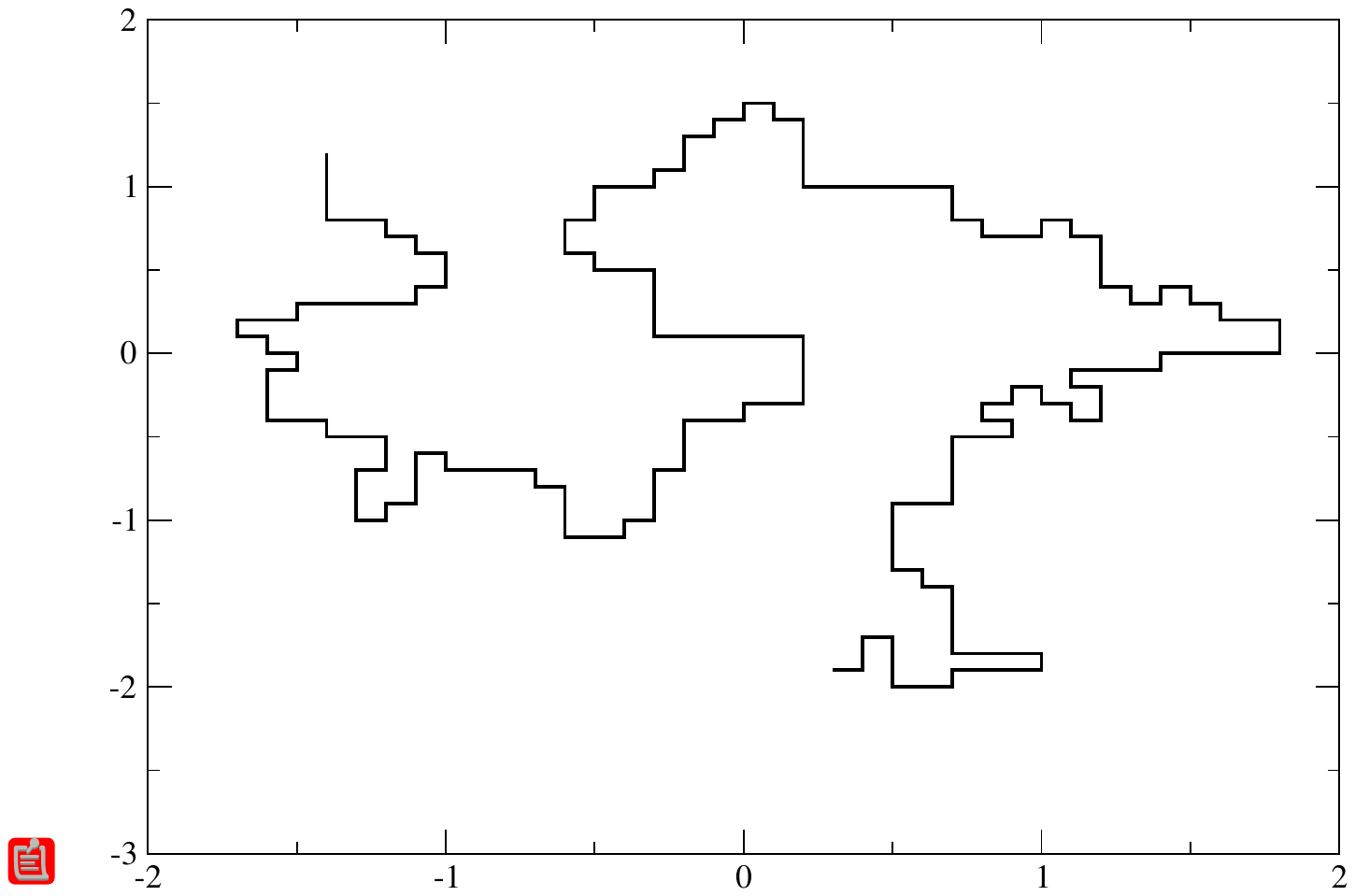}
\includegraphics[width=0.45\linewidth]{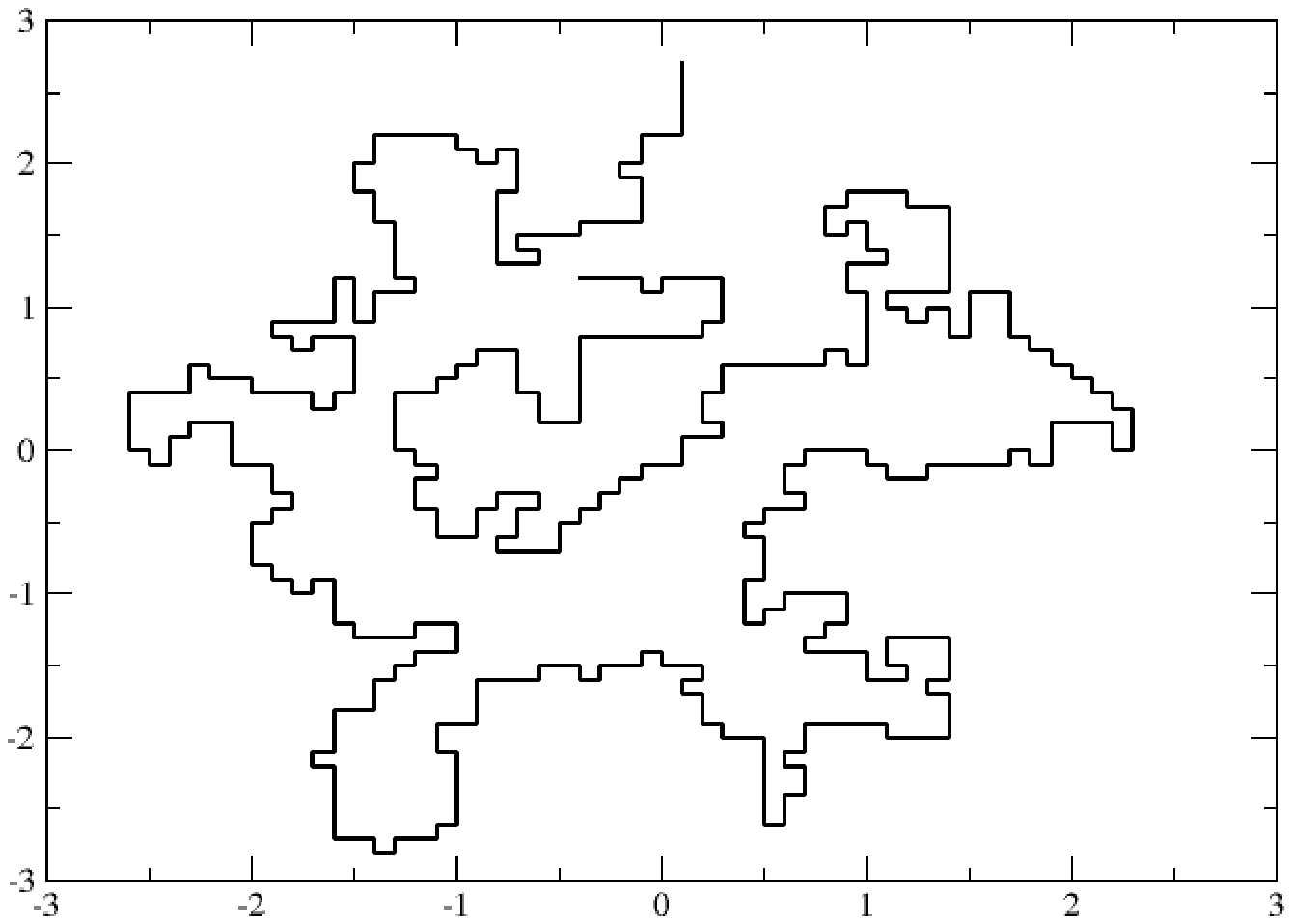}
\caption{Best configurations obtained with simulated annealing for 
different values of $L=5,10,20,50$. The simulations are obtained with a
lattice polymer and using the pivot algorithm \cite{Madras:1988}.}
\label{fig:SA}
\end{figure}

Our numerics thus suggest there are 4 sharp transitions of shape: C-shape
to cross near $L = 3.2$, cross to S-shape near $L = 5.8$, S-shape to
star near $L=8.7$ and finally a transition from the star network to the
spiral near $L = 110$.  But the star only slightly out-performs these
curves, so it is possible that in fact there is a smooth evolution of
curves as optimal networks.  An alternative numerical approach is to
seek the overall optimal network, via simulated annealing for example.
This is computationally difficult, but some preliminary results for
optimal curves are shown in Fig.~\ref{fig:SA}, and are roughly
consistent with our qualitative summary.

\section{The minimum distance to the center}

For our second variant, closer to the general problem, we seek the
network of length $L$ that minimizes the average time to reach a
designated ``center" location. This was for example discussed in
\cite{Bejan:1996,Bejan:1998} where the case of street networks was
considered and where the optimal tree was found. This problem in the
context of a single line bus was also considered in \cite{Okabe:1992}
(and references therein). With respect to transportation systems such
as subways or trains, this is obviously a crude simplification as we
are in general interested in reaching many other stations and not a
single location. As we will see in the rest of the paper, our results
suggest that this simplified problem perhaps captures the essence of
the general problem and might constitute a useful toy model where
analytical calculations are feasible.  As in our other variants we
envisage each position on the network as a ``station" where the subway
is accessible.

Taking as before an isotropic density $\rho( \ )$ and the origin as
``center", we seek the ``optimal" network that minimizes
\begin{align}
\overline{\tau}= \int\tau(z,0)  \rho(z) dz
\end{align}
where $\tau(z,0)$ is the minimum time to go from the point $z$ to the origin $0$.
The optimal network will depend on the parameter 
$S = v_s/v_w > 1$, where $v_s$ is the speed within the network, and $v_w$ the speed outside.

In order to simplify analytical calculations we assume in this variant
that the paths from the points to the network can be made only along
circular ($r=\mathrm{const.}$) or radial ($\theta=\mathrm{const.}$)
lines.  It remains true that the overall optimal network must be a
path or tree. Indeed, if there is a circuit in the network, there is at
least one point such that starting in either direction takes the same
time to get to the center. One can then remove a small interval from
that point, and re-attach elsewhere to get a better network.

However we will consider only simple shapes for the
network, starting from the star network and then adding a ring to
it. As we will see below, for these structures we can develop simple
analytical calculations and observe important phenomena such as a
topological transition.

\subsection{Star network}

We again consider the star network, having $n_b$ branches of lengths
$r^*$, outward from the origin, evenly spaced with angle $2\pi/n_b$
spacing (see Fig.~\ref{fig:mainshapes}(d)). So $L = n_b r^*$ and (for
given $L$) $n_b$ is a free parameter to be optimized over. By isotropy we can write
\[ \rho(x,y) = \rho(r) \mbox{ for } r^2 = x^2 + y^2 . \]
Again by isotropy, the average time $\overline{\tau}$ to the center is such that
\begin{align}
 \frac{1}{2n_b}\overline{\tau}=&\int_0^{\pi/n_b}\mathrm{d}\theta
\Big[ \int_0^{r^*}\mathrm{d}rr\rho(r)(\frac{\theta
  r}{v_w}+\frac{r}{v_s})\nonumber \\
&+\int_{r^*}^R \mathrm{d}rr\rho(r)(\frac{r-r^*}{v_w}+\frac{\theta
  r^*}{v_w}+\frac{r^*}{v_s}) \Big]  . \label{eq:int2}
\end{align}
In the following we will consider the uniform density on a disc,
and an exponentially decreasing density.

\subsubsection{Uniform density}

\begin{figure}[ht!]
\includegraphics[width=0.45\linewidth]{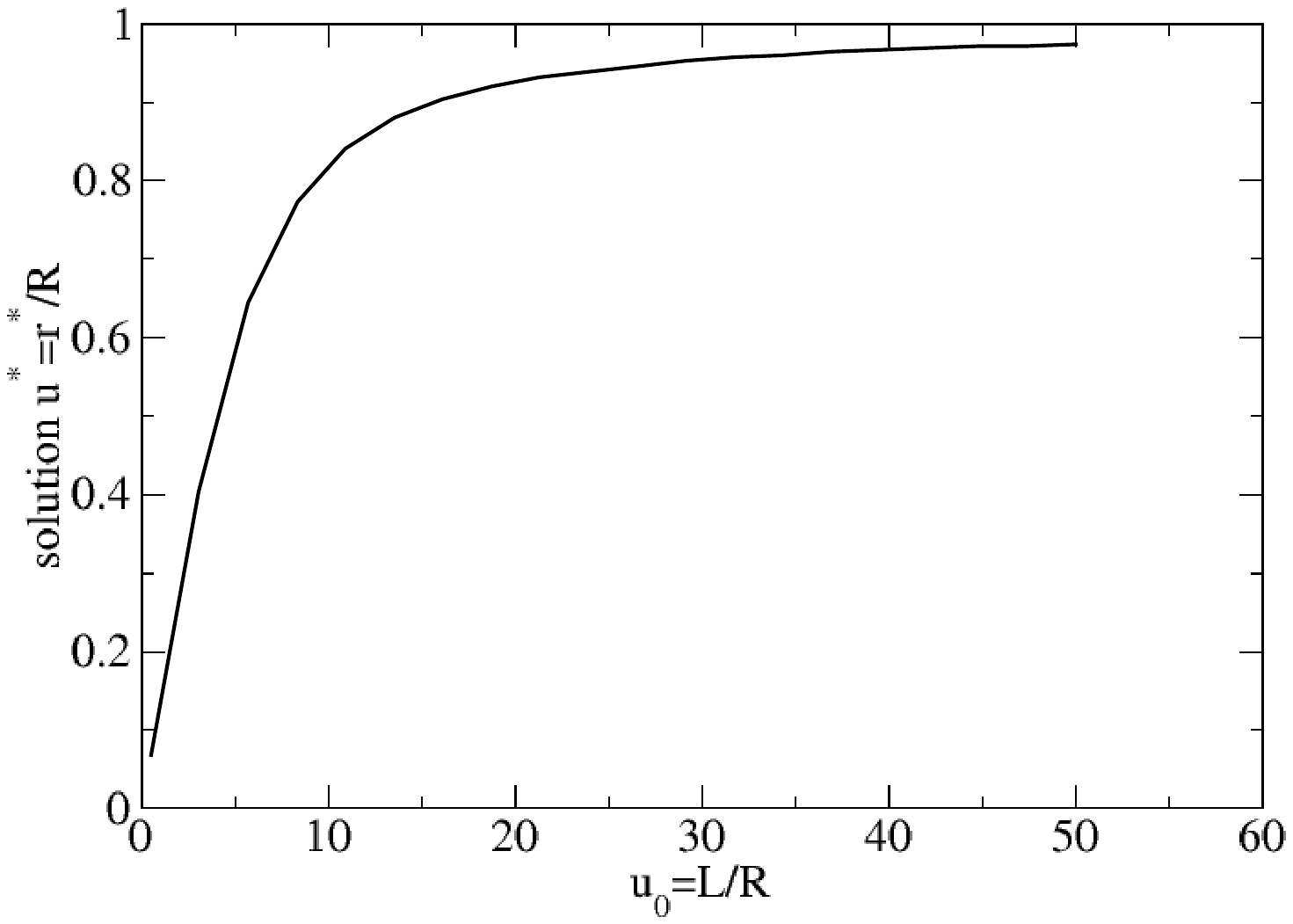}
\includegraphics[width=0.45\linewidth]{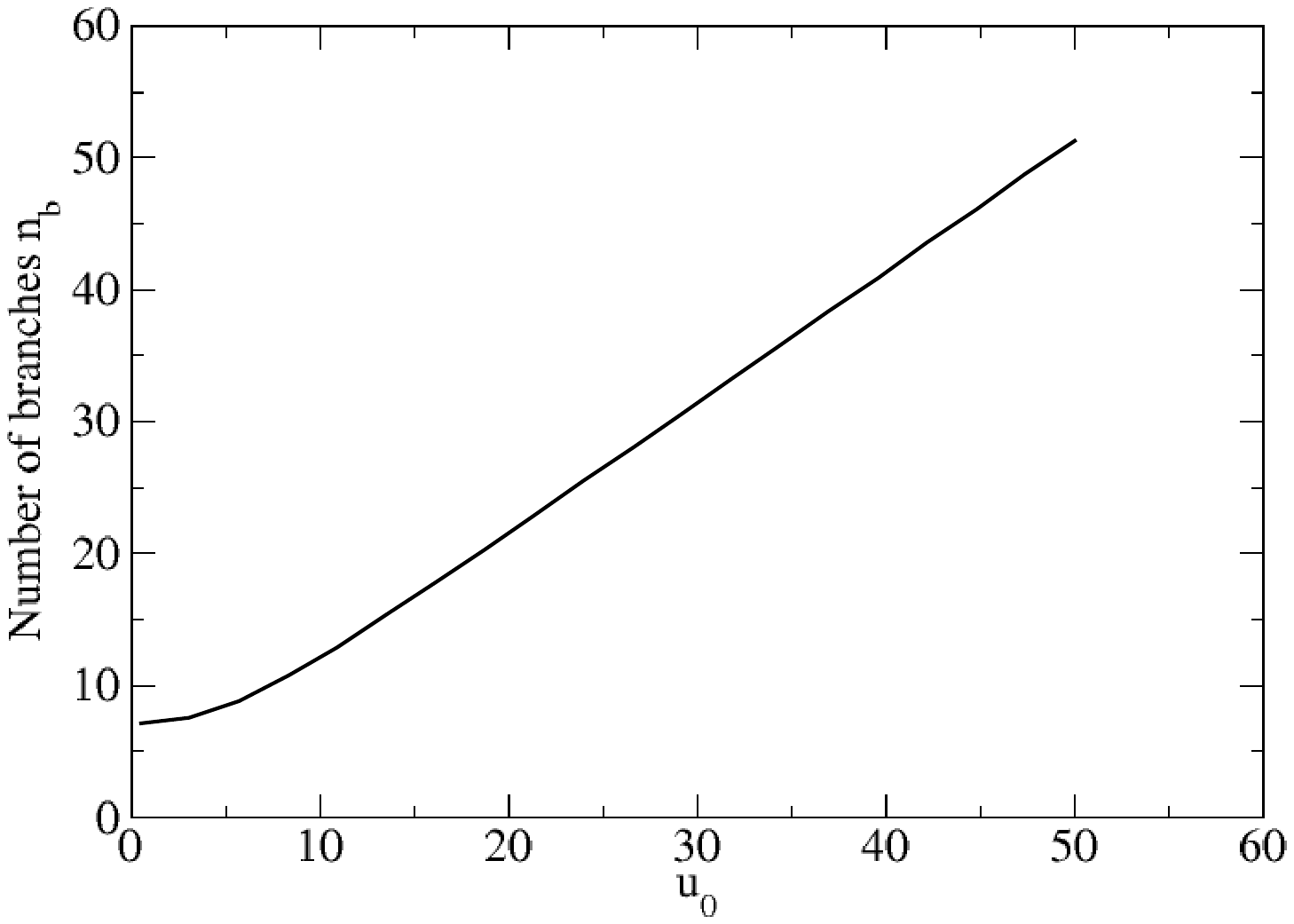}\\
\includegraphics[width=0.45\linewidth]{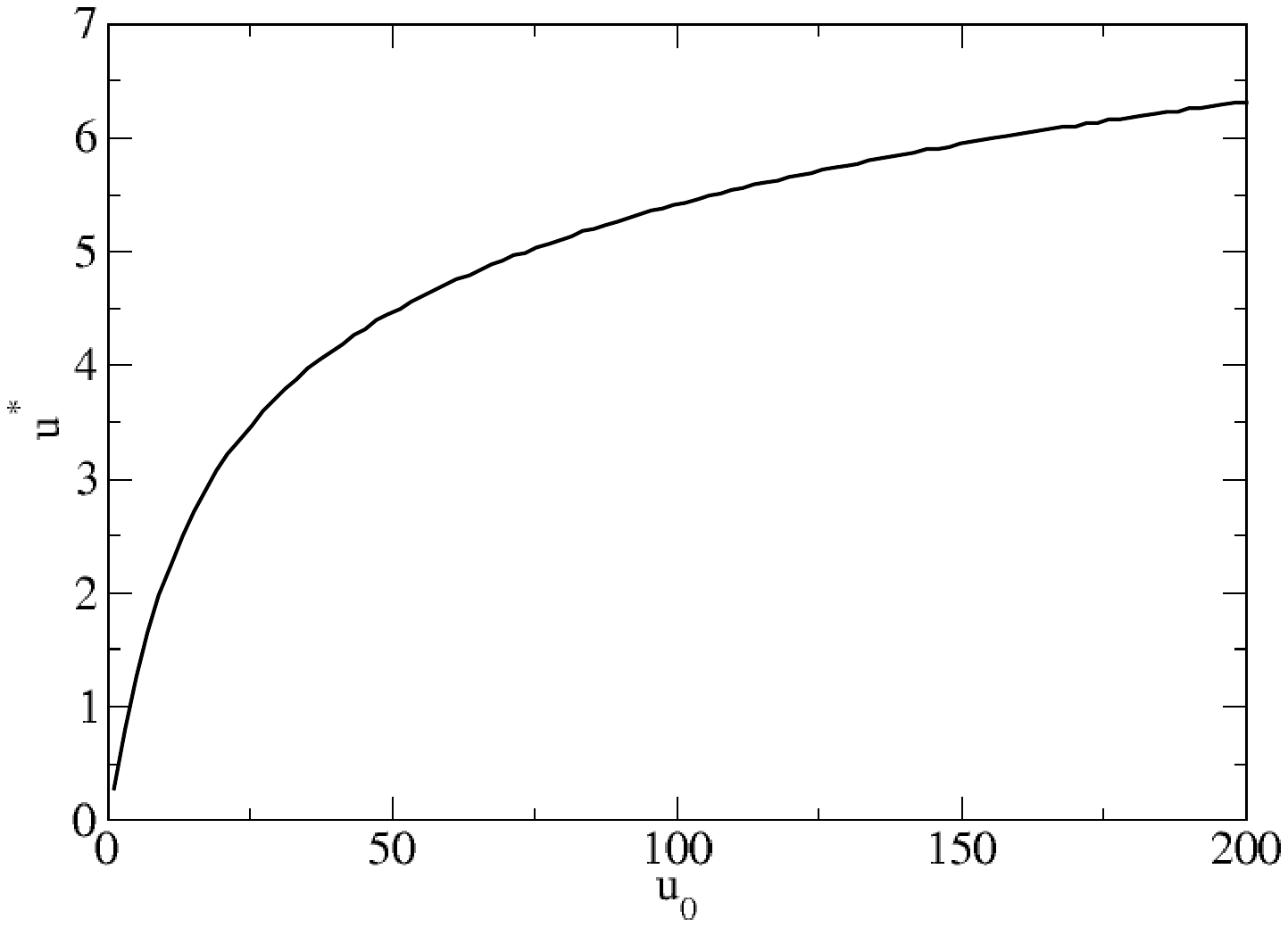}
\includegraphics[width=0.45\linewidth]{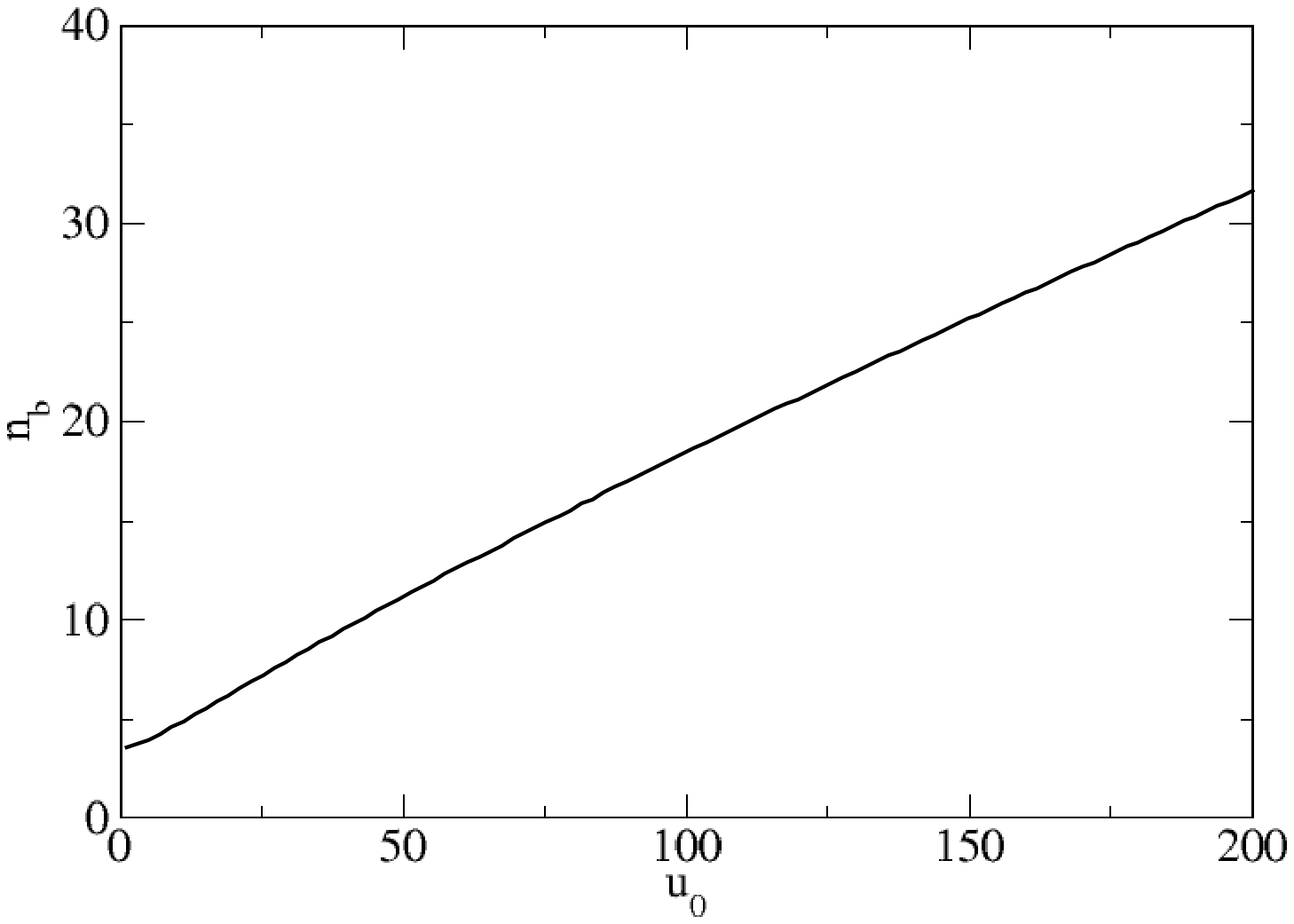}
\caption{We show here the  result of the minimization of the average  time for
  the star network with parameters $n_b$ and $r^*$ with the constraint
  $L=n_br^*$. On the left we show the length of branches $r^*$ versus
  $u_0=L/R$ and on the right the number of
  branches $n_b$ versus $u_0$ (here $\eta=1/8$). The top row
  corresponds to the uniform density case and the bottom one to the
  exponential density.}
\label{fig:star}
\end{figure}

Here we take the uniform density
$\rho(r) =\rho_0 = 1/(\pi R^2)$ 
on a disk of radius $R$. 
Without the network the average time to reach the center is
$\tau_0 =  2R/(3v_w)$. Write $\eta =1/S =  v_w/v_s$, and assume that taking the
 subway is always better than walking directly to the center, which is the condition that $\eta\leq 1-\pi/n_b$.
Write $u^*=r^*/R$ (the branch length relative to city radius) and 
$u_0=L/R$ (network length relative to city radius) and
$\chi = \pi/n_b$.
Evaluating the integrals in (\ref{eq:int2}), the average time $\overline{\tau}$ 
to reach the center via subway satisfies
\begin{align}
\nonumber
\frac{\overline{\tau}}{\tau_0}=&\frac{u^{*3}}{2}[-\frac{\chi}{2}-\eta+1]\\
&-\frac{3}{2}u^*[-\frac{\chi}{2}-\eta+1]+1 .
\end{align}
 For given $L$ we want to optimize over the free parameter $r^*$, that is over $u^*$.
 From $L=n_b r^*$ we obtain $\chi=\pi u^*/u_0$ and then the average time as a function of
$u^*$ reads
\begin{align}
\frac{\overline{\tau}}{\tau_0}=\frac{1}{2}(u^{*3}-3u^*)[-\eta+1-\pi\frac{u^*}{u_0}]   +1 .
\end{align}
Minimizing  this quantity over $u^*$ leads to a
polynomial of degree 3, and 
the behavior of the solution is shown numerically in
 Fig.~\ref{fig:star}.
As is intuitively obvious, for large $u_0 = L/R$ it is optimal to use
roughly $u_0$ branches of length almost $R$; more precisely for
$u_0 \gg 1$ we obtain
\begin{align}
\begin{cases}
u^*&=1-\frac{\pi}{3(1-\eta)u_0}+{\cal O}(\frac{1}{u_0^2})\\
n_b&=u_0+{\cal O}(1) .
\end{cases}
\end{align}
Perhaps less obvious is the initial behavior over $0 \le u_0 \le 10$, where
the length of
branches is increasing faster than their number. In other words we
first observe a radial growth and then an increase of the number of
branches.

\subsubsection{Exponential density}

In this case, we take  the density of the radial component to be 
$\rho_0 r \mathrm{exp}(-r/r_0)$ on the infinite plane. 
Without the network the average time to reach the center is
$\tau_0 = 2r_0/v_w$. Evaluating the integrals in (\ref{eq:int2}), the average time $\overline{\tau}$ 
to reach the center via subway satisfies
\begin{align}
\nonumber
\frac{\overline{\tau}}{\tau_0}&=
\frac{\pi u^*}{2u_0}+\eta\\
&-\mathrm{e}^{-u^*}\left(\frac{\pi
  u^*}{u_0}+\eta-1\right)\left(\frac{u^*}{2}+1\right)
\end{align}
where $u^* = r^*/r_0$ and $u_0 = L/r_0$.
We can plot this function and look numerically for the
minimum. The results are shown in Fig.~\ref{fig:star} (bottom).

We observe here that for large resources ($u_0\gg 1$) the number
of branches scales as $n_b\sim au_0$ with $a\approx 0.1$
and the solution $u^*$ seems to converge slowly to some value that
depends on $\eta$.

\subsection{Loop and branches}

\begin{figure}[ht!]
\includegraphics[width=0.6\linewidth]{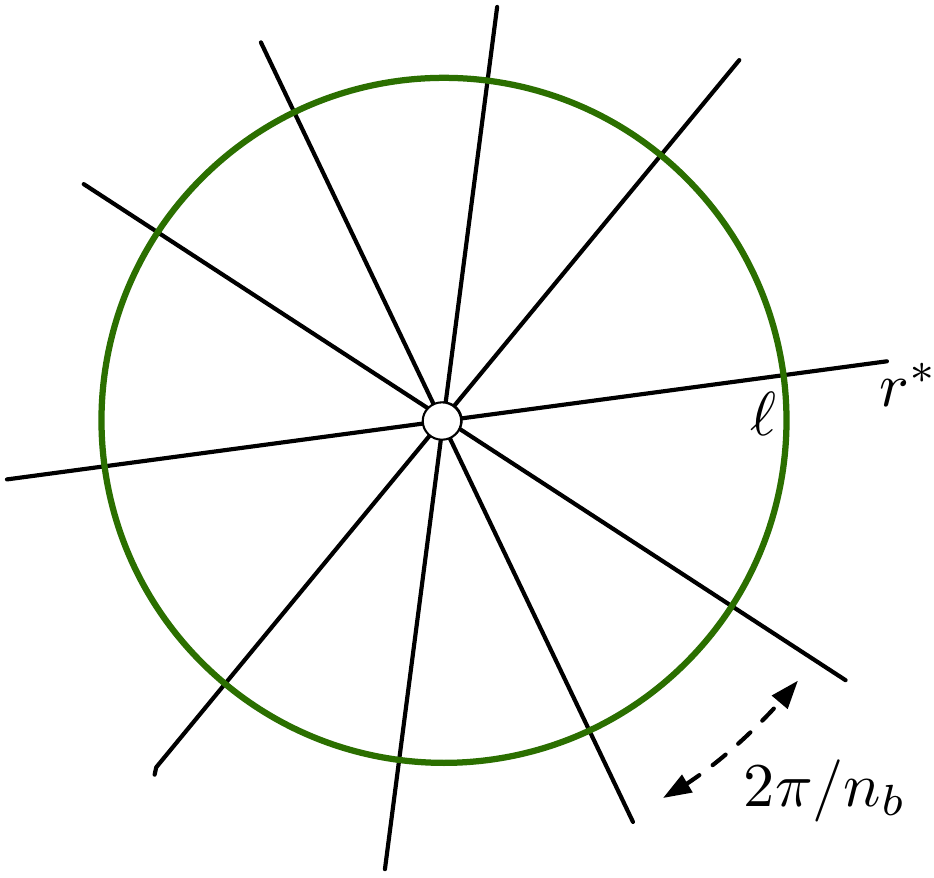}
\caption{Schematic of the star network combined with a ring. We have now
  three parameters: the number
  of branches $n_b$, the length $r^*$ of the branches, and the radius
  $\ell$ of the ring.}
\label{fig:all}
\end{figure}

We now consider a more interesting case where we have
$n_b$ branches of length $r^*$ and a ring of radius $\ell$ (see
Fig.~\ref{fig:all}). 
This case is essentially motivated by subway networks
that seem to display this type of structure when they are large enough
\cite{Roth:2012}. 
We have 3 parameters: $n_b$, $\ell$ and $r^*$, where  $\ell\leq r^*$ for connectivity.  The total
length of the network is
\begin{align}
L=2\pi\ell+n_br^* .
\end{align}
This network enables us to study  the relative contributions of
loop and branches to our goal of minimizing the average
time to go to the center. This case is analytically difficult, so we study
it numerically with a simulation. 

\begin{figure}[ht!]
\includegraphics[width=0.8\linewidth]{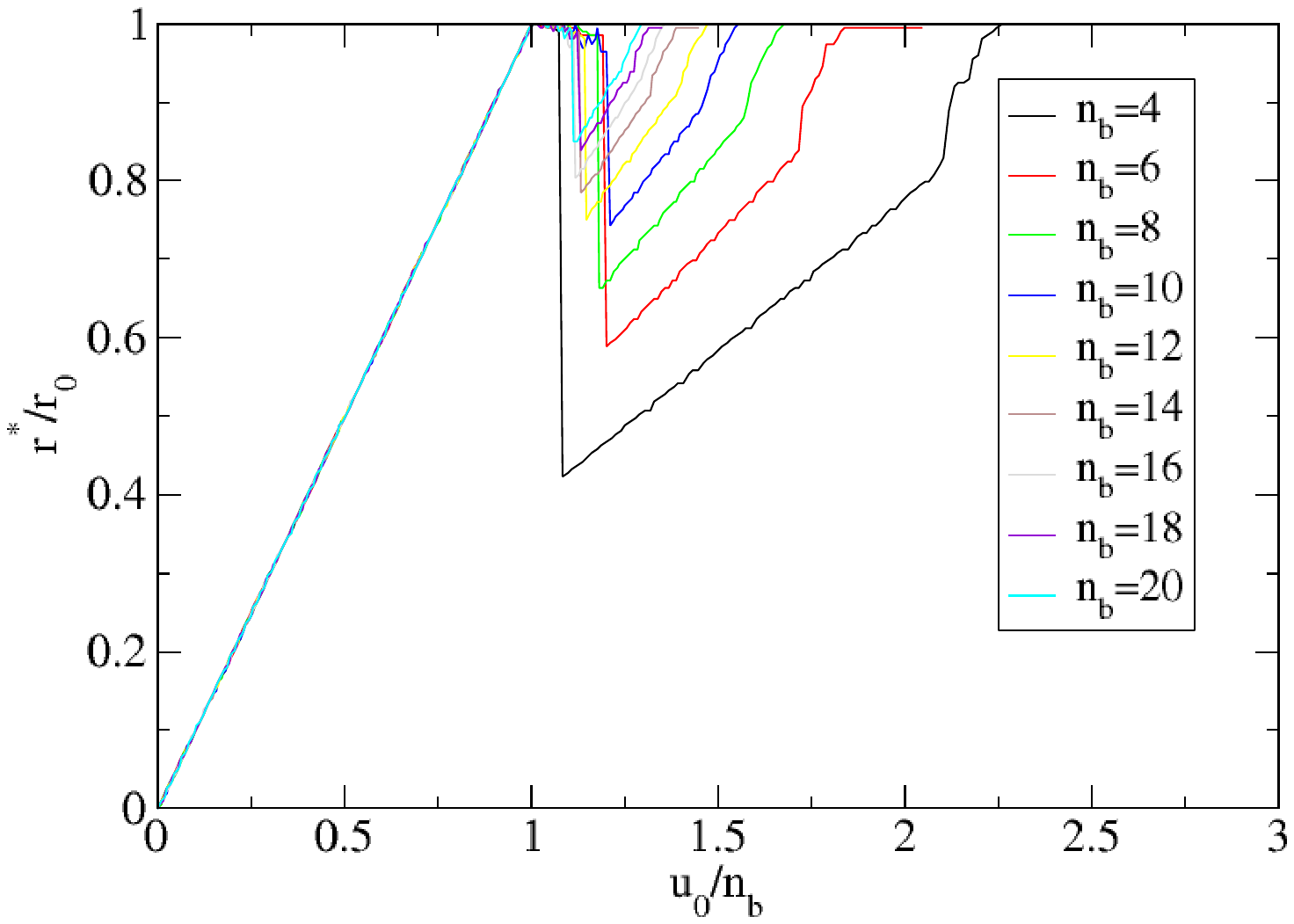}
\includegraphics[width=0.8\linewidth]{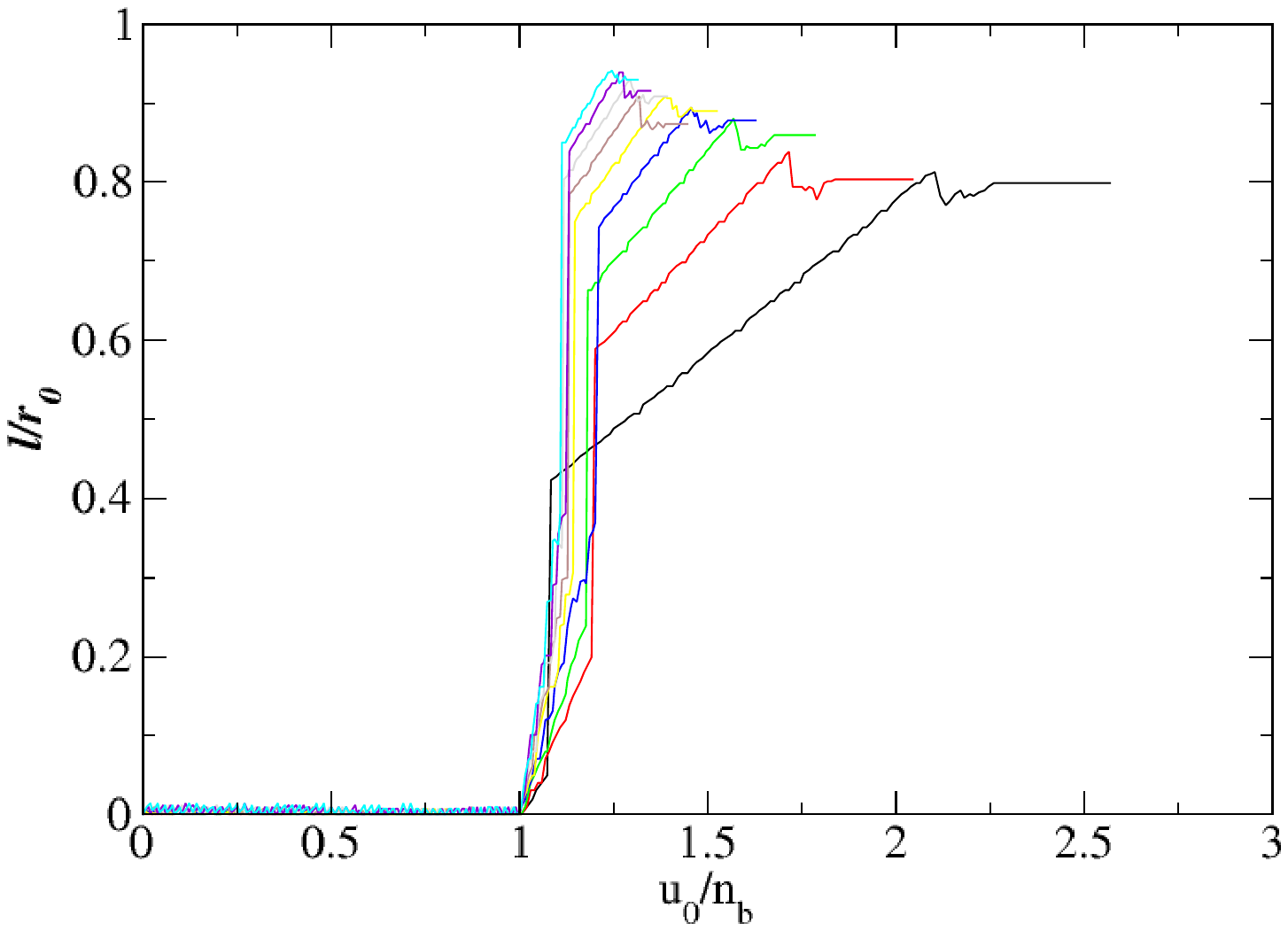}
\caption{Uniform density: results for the star+loop network for
  different values of $n_b$. (Top) Normalized radius of the
  ring. (Bottom) Normalized length of branches. We normalized $u_0$
  by $n_b$ in order to get the same ``transition" point at
  $u_0/n_b=1$. These results are obtained for $\eta=1/8$.}
\label{fig:all2}
\end{figure}

We first consider the uniform distribution on a disc of radius $r_0$,
and again choose $\eta=1/8$ (which corresponds to the reasonable
values $v_w\approx 5$km/h and $v_s\approx 40$km/h). We first study the case
where $n_b$ is fixed and where we optimize the network over $r^*$ and
$\ell$. In Fig.~\ref{fig:all2}, we show the results for the optimal
value of $r^*$ and $\ell$ versus $u_0=L/r_0$ normalized by $n_b$.
We observe that when resources are growing from $0$, we have only a
radial network ($\ell=0$). At $u_0/n_b=1$ we have a ``transition" point
where a loop appears. This means that until $u_0=n_b$ all the
available resource is converted into the radial structure. When the
radial structure is at its maximum ($r^*=r_0$) we observe the
appearance of a loop whose diameter is then increasing with $u_0$.

We note that even if both the number of branches and the size of the
loop undergo a discontinuous transition, the average minimum time
displays a smooth behavior. Also, if we increase further the total
length $L$, it will result in a larger number of branches $n_b$.

\begin{figure}[ht!]
\includegraphics[width=0.8\linewidth]{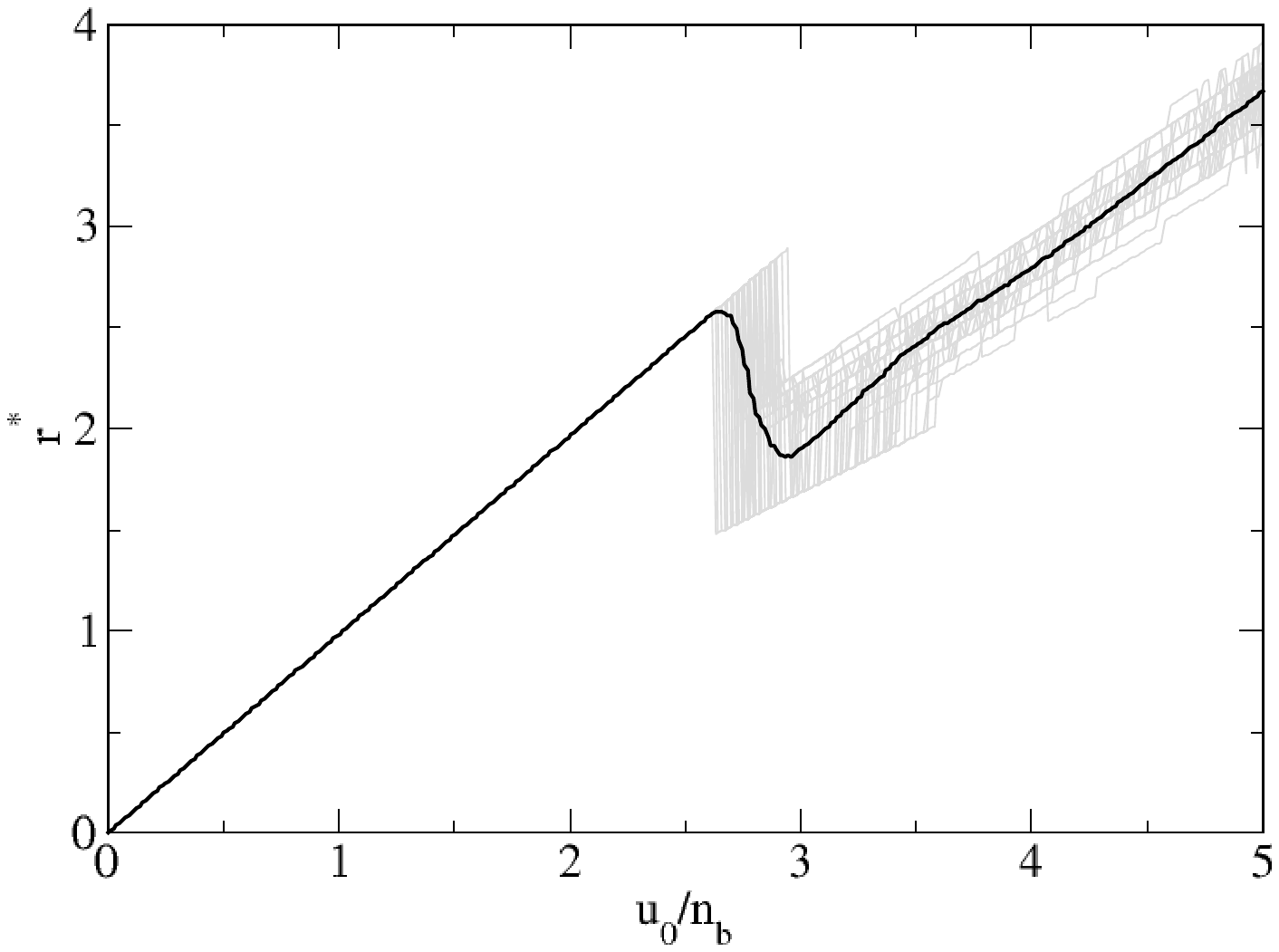}
\includegraphics[width=0.8\linewidth]{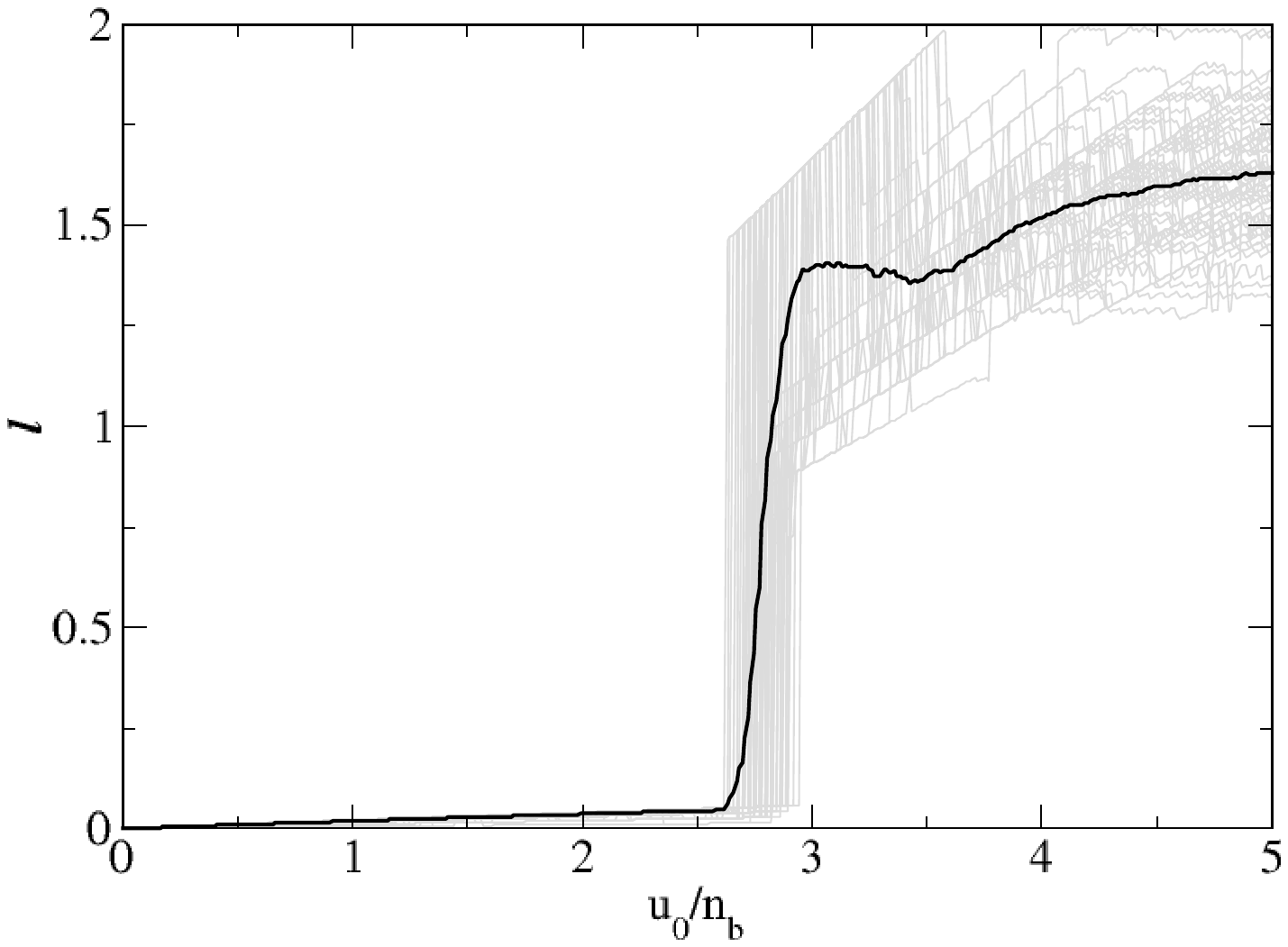}
\caption{Gaussian density ($\sigma=1$): results for the star+loop
  network for $n_b=8$. (Top) Size $r^*$ of the branches. (Bottom) Radius of the
  ring. These results are averaged over $100$ configurations and are
  obtained for $\eta=1/8$. The grey curves represent the results for
  each configuration.}
\label{fig:gauss}
\end{figure}

In the uniform density case, the domain is finite (a disk of radius
$R$) and at fixed value of $n_b$, there is therefore a maximum value
of $L_{max}=n_bR+2\pi R$. For larger values of $L$, the optimal network
will increase its number of branches $n_b$. It is different in the
Gaussian disorder case: the domain is infinite and there is no obstacle
to have a fixed value of $n_b$ with size $r^*$ growing indefinitely with $L$. We
can thus expect some differences with the uniform density case. We
repeated the calculations above for $100$ configurations and the
average together with the results for each configuration are shown in 
Fig.~\ref{fig:gauss}.
We still observe the different regimes separated by an abrupt
transition: the first regime where the size of branches grows with $L$
and the second regime where there is a ring whose size grows very
slowly with $L$. In contrast with the uniform density case, the transition takes place for
a value $u_0/n_b$ that fluctuates in the range $[2.5,3.0]$.

\section{The general model}

As discussed in the introduction, we will now consider the ``general"
setting of routes between arbitrary points $z_1$ and $z_2$.  The route
can either go straight (speed $v_w$) from $z_1$ to $z_2$ without using
the network, or go $z_1 \to A_1 \to A_2 \to z_2$, where $A_1$ and
$A_2$ are ``stations" (any points on the network) and where travel
from $A_1$ to $A_2$ is within the network at speed $v_s$, and the other
journey segments are straight at speed $v_w$.  In a companion paper
\cite{AB2} we study the more realistic model where the route is
optimized over choice of $A_1$ and $A_2$, but here we take each $A_i$
as the closest station to $z_i$.  Unlike previous cases, the overall
optimal network is not necessarily a tree.


\subsection{Various shapes}

\begin{figure}[ht!]
\includegraphics[width=0.8\linewidth]{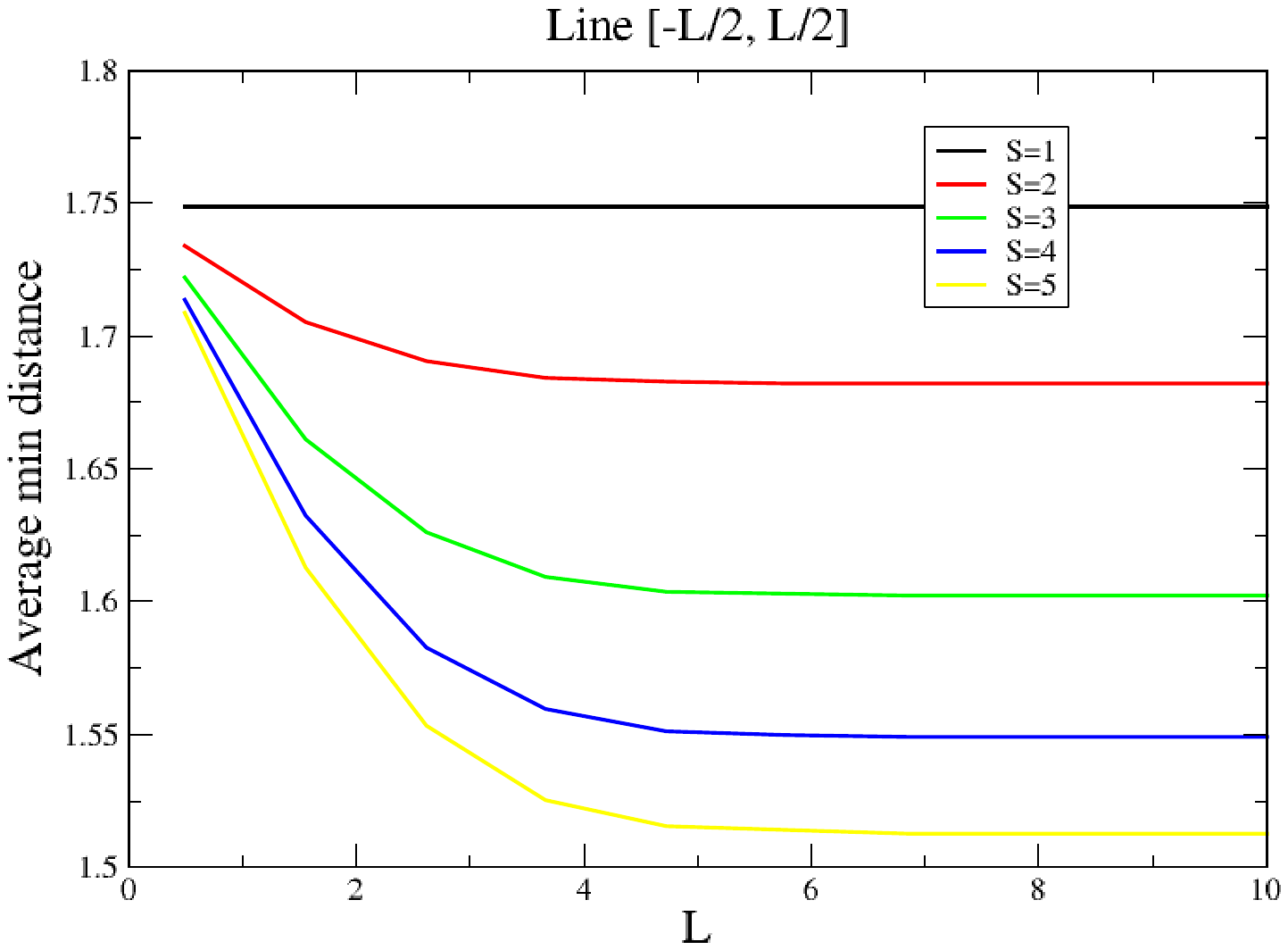}
\includegraphics[width=0.8\linewidth]{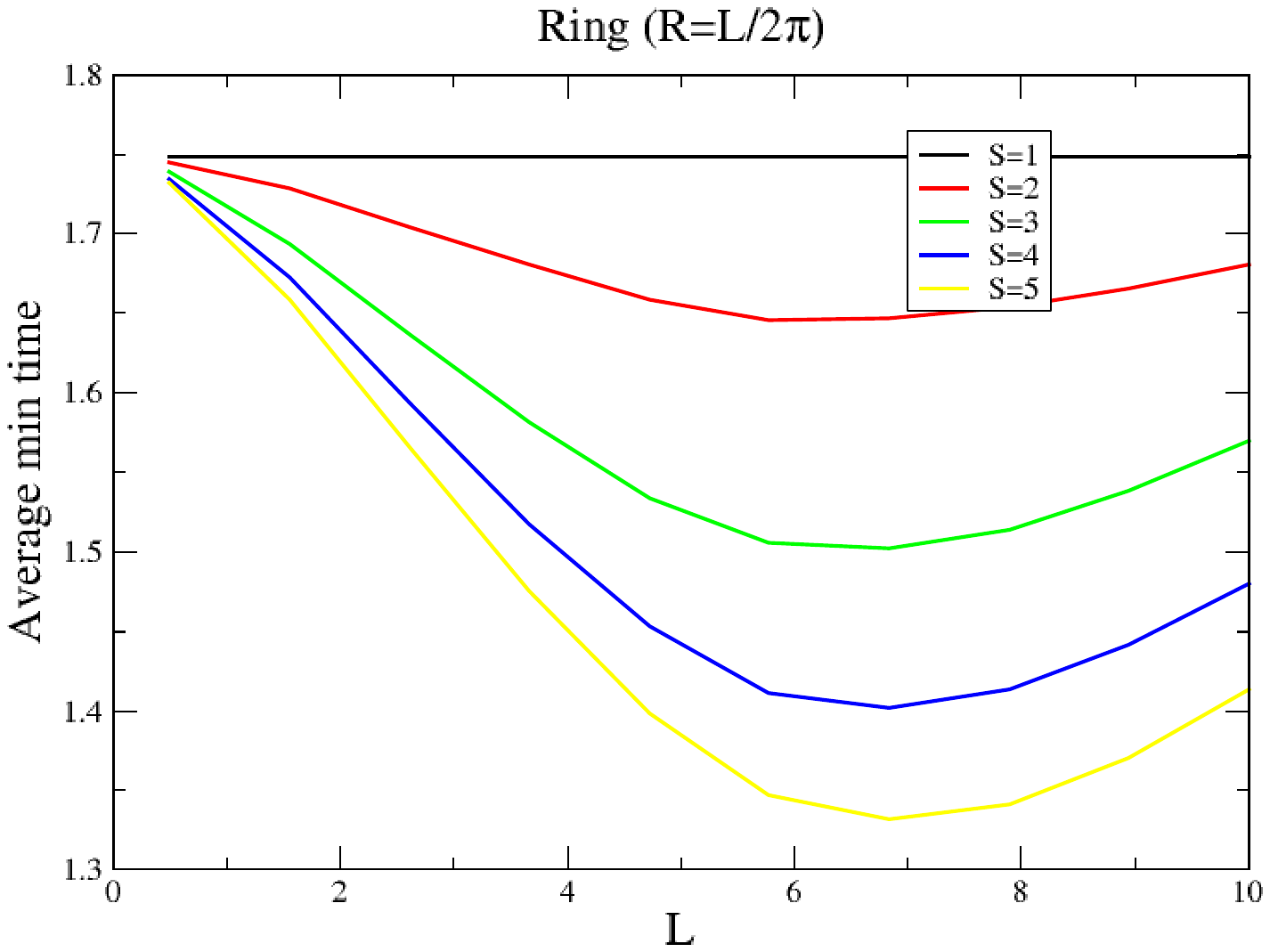}
\caption{Average journey time in the Gaussian case for different
  values of $S$ and for (Top) a line and (Bottom) a ring.}
\label{fig:gpline}
\end{figure}

We present results for various simple shapes, for the standard Gaussian density.
We start with the line segment and the ring and the results for
different values of $S$ are
shown on Fig.~\ref{fig:gpline}.
For these two shapes, we observe the same behavior as in the 
time to center problem: for the line there is a quick saturation to a
constant, and for the ring there is a minimum at $L\approx 2\pi\sigma$.

\begin{figure}[ht!]
\includegraphics[width=0.9\linewidth]{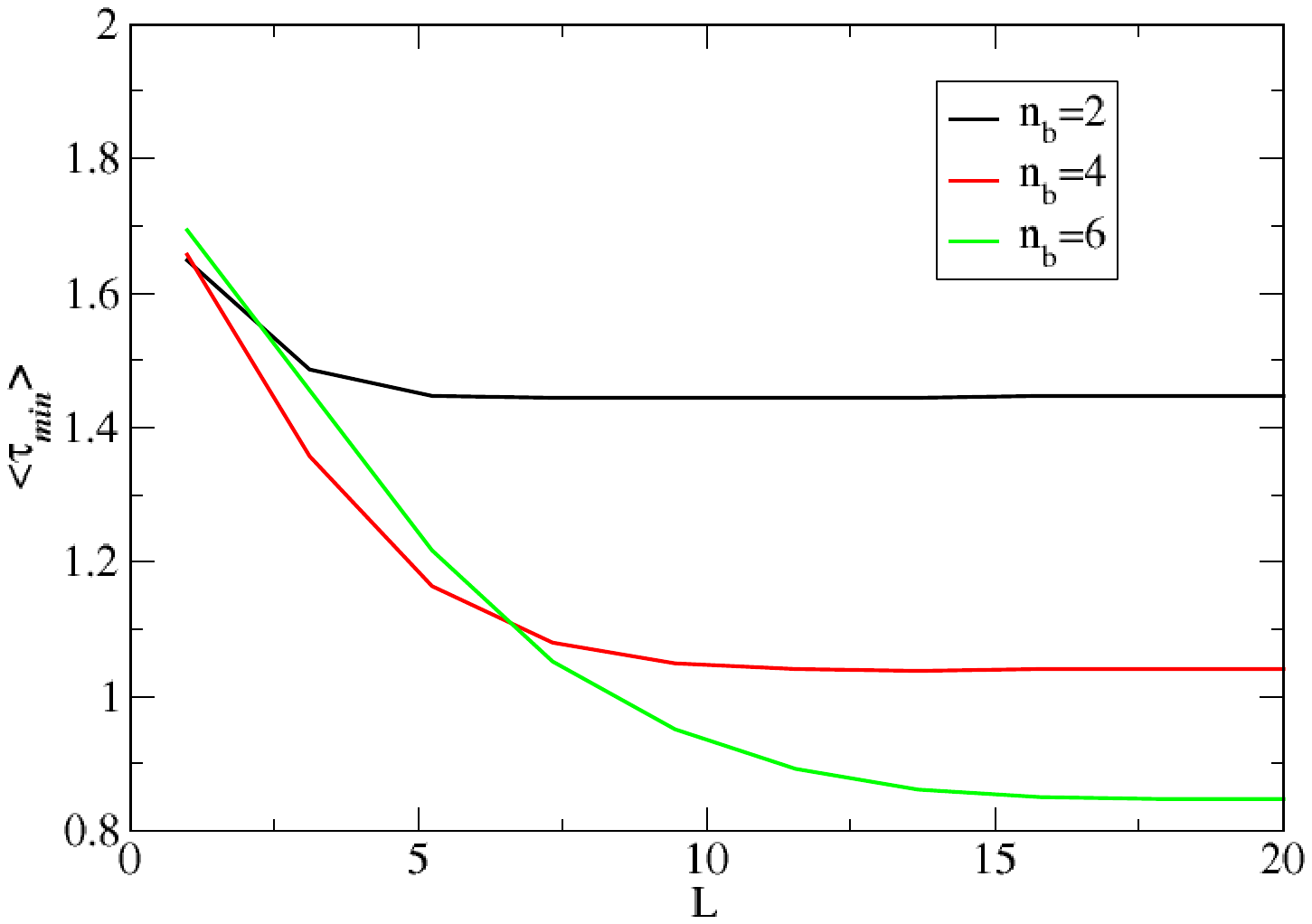}
\caption{Average journey time (in the Gaussian case) for star networks
  with different number of branches ($n_b=2,4,6$).}
\label{fig:gpcross}
\end{figure}

We also consider the case of the star network with $n_b$ branches and
the result is shown on Fig.~\ref{fig:gpcross}.
We observe that for $0<L\lesssim 1.15$ the line is optimal. For
$1.15\lesssim L\lesssim 6.6$, the cross $n_b=4$ is the optimal
choice, while for $L\gtrsim 6.6$, the solution $n_b=6$ is better. Very likely
we will have (as in the previous case of the average time to the
center) an optimal network with $n_b\propto L$.

\subsection{Loop and branches: Scaling of the average  time}

\begin{figure}[ht!]
\includegraphics[width=0.9\linewidth]{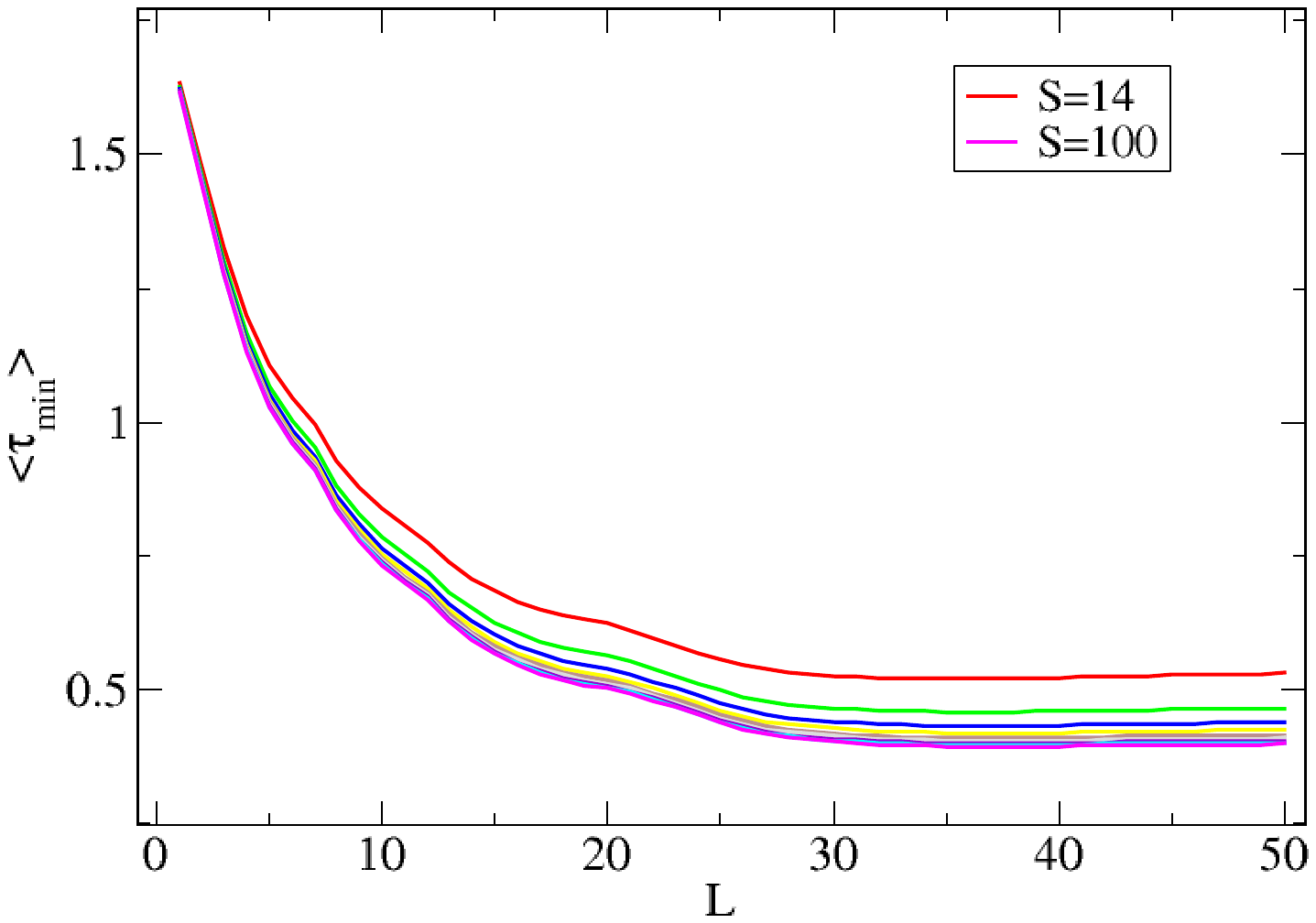}
\caption{Average  time (Gaussian density) for $n_b$
  branches plus a loop for values of $S$ from $14$ to $100$.}
\label{fig:allS}
\end{figure}

\begin{figure}[ht!]
\includegraphics[width=0.9\linewidth]{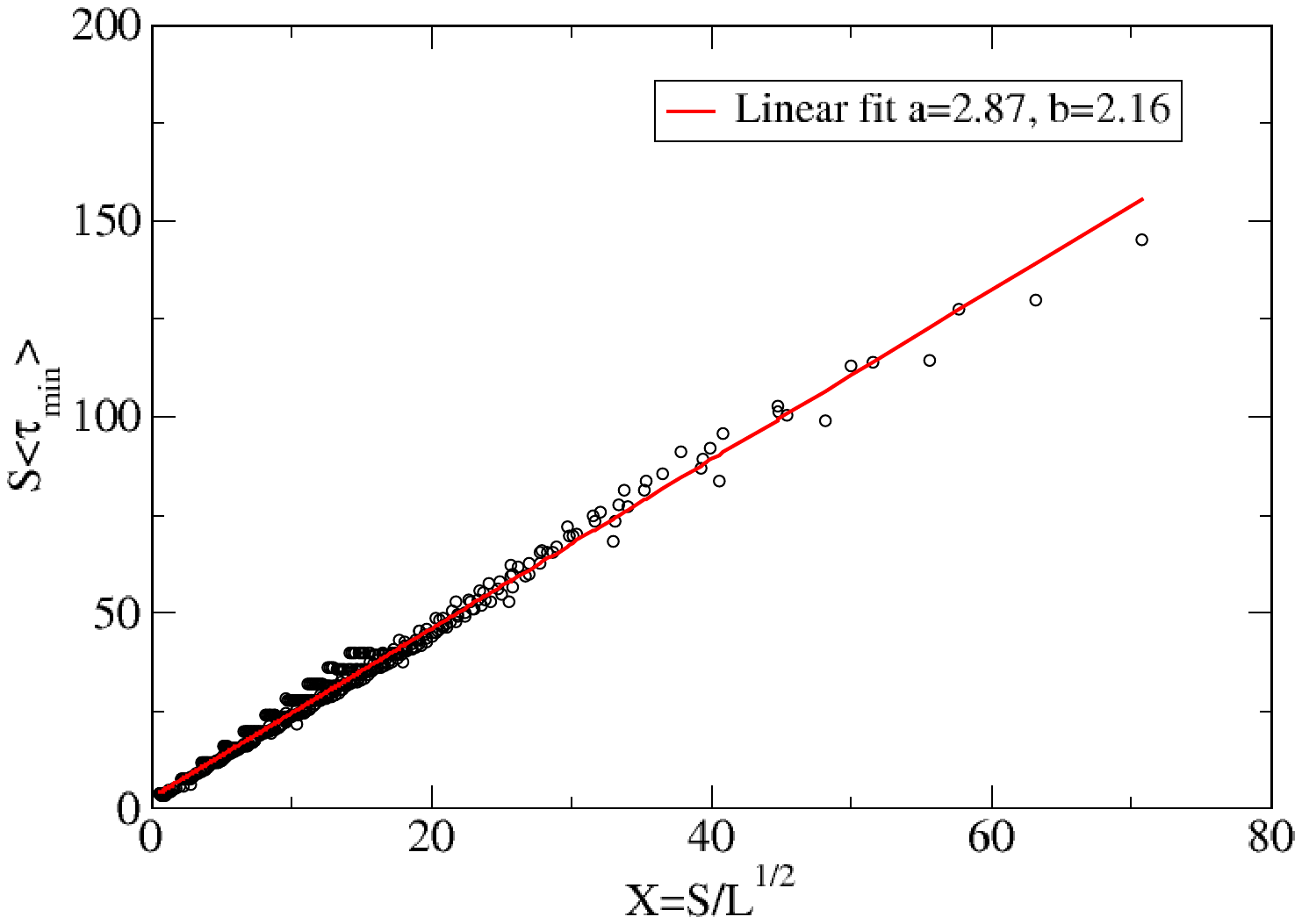}
\caption{Rescaled average  time $S\overline{\tau}_{min}$ versus the rescaled variable
  $S/\sqrt{L}$.}
\label{fig:collapse}
\end{figure}

We focus here on the case where the network  is made of $n_b$
branches of length $r^*$ and a loop of radius $\ell$. So
$
L=n_br^*+2\pi\ell . $
We take  $n_b$ and $\ell$  as the 2 free parameters over which we 
 will minimize the average time.  The optimized average time $\overline{\tau}_{min}$
 for different values of $S$ is shown  in Fig.~\ref{fig:allS}.
Naively one expects that this quantity
behaves as
\begin{align}
\overline{\tau}_{min}=\frac{a}{\sqrt{L}}+\frac{b}{S}
\label{eq:collapse}
\end{align}
where the first term of the r.h.s. corresponds to the average 
distance to the network and which we expect to scale as
$1/\sqrt{L}$. The second term corresponds to the shortest path
distance within the network. In principle $a$ and $b$ could depend on
$L$. If we assume this form to be correct then
$S\overline{\tau}_{min}$ versus $X=S/\sqrt{L}$ should be a straight
line. We tested this assumption on the data from Fig.~\ref{fig:allS}
and the result is shown in Fig.~\ref{fig:collapse}.
This good collapse (except for deviations observed for large values of
$S/\sqrt{L}$ supports the assumption (\ref{eq:collapse}).

\subsection{Loop and branches: A topological transition}

\begin{figure}[ht!]
\includegraphics[width=0.9\linewidth]{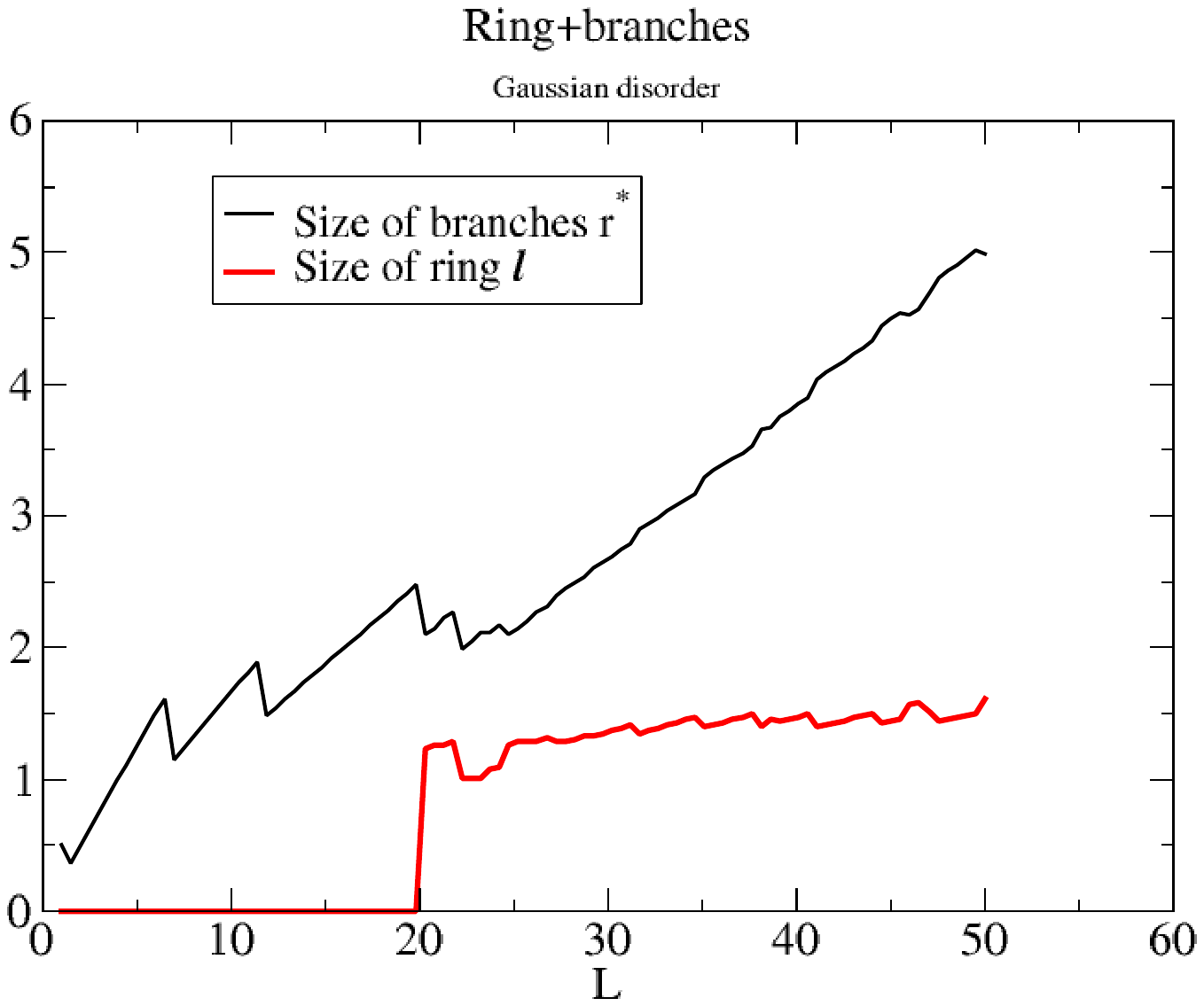}
\caption{Optimal length of branches and optimal size of the loop
 versus the total length $L$ (for gaussian disorder): typical shape for $r^*$ and $\ell$. As
 in the case of the quickest path to the origin, we observe a sharp transition for
  $L\approx 20\sigma$ where a loop of radius $\ell\approx \sigma$
  appears (here $\eta=1/2$).}
\label{fig:gptransit}
\end{figure}

Still in the branches and loop model, we observe  (Fig.~\ref{fig:gptransit})   a
transition:
$r^*$ grows almost steadily with $\ell=0$,
until a transition value at $L_c\approx 20$ where the ring appears.
Although the structure changes abruptly it is interesting to note that
there is no discontinuity in the average minimum time.  The size of
the ring stays stable with $\ell\approx\sigma$ (the Gaussian s.d.). 
Naively, we can say that the ring appears when the
branches can have a length $r^*\approx\sigma$ and a loop of size $\ell\approx \sigma$ which gives the condition
$L_c\approx n_b\sigma+2\pi\sigma$. At the transition, we observe that
we have $n_b\approx 8$ which then gives $L_c\approx 14$, not too far from the value $L_c=20$ observed here. 

The
quantity $L_c$ is independent from $S$ which is expected, as it is
essentially controlled by the topology of the network.  We note that this
transition was already observed in the previous case ``minimum distance to the
center". As $L$ increases the optimal number of branches
grows roughly from $2$ to $10$. The picture that emerges here is
consistent with the empirical study \cite{Roth:2012}: we observe a
ring around the ``core" of the city and then branches radiating from
it. More generally, these results suggest that the distance to
center problem is a reasonably good proxy for the more general
problem.


\section{Discussion}


Algorithmic aspects of network design questions similar to ours have
been studied within computational geometry (e.g. \cite{Okabe:1992}
chapter 9) and ``location science" (e.g. \cite{Laporte:2015}).  But
our specific question -- optimal network topologies as a function of
population distribution and network length -- has apparently not been
explicitly addressed. Although real-world networks are probably not optimal and result from
the superimposition of many different factors, understanding theoretically optimal 
 networks could give some information about the actual
structure observed in many cases. For example, it could help us to
understand the seemingly universal structure displayed by very large subway
networks \cite{Roth:2012}. 

Even for simple models, optimizing over all possible topologies is difficult,  so 
 we investigated only various simple shapes. We provided general
arguments and analytical calculations in simple cases and most of our
analysis is numerical. In general we expect an evolution of the shape
of optimal networks when $L$ increases, with the possible existence of
sharp transitions between different shapes. Although we weren't able
to prove this in general, we observed such transitions in simple cases such as
branches and a loop: in the case of Gaussian (variance $\sigma^2$) population density, starting from a
small value of $L$ the branches first grow smoothly, then suddenly for a value
$L_c$ we observe the appearance of a loop of size $\ell\sim\sigma$.
This transition also exists in the case of uniform density.

It would be interesting to see these transitions of overall optimal networks
obtained numerically, and this might be feasible in some cases with
a simulated annealing type of algorithm. In any case these problems
suggest theoretical questions and practical applications which certainly
deserve further studies.




\end{document}